\documentclass[preprint,12pt]{elsarticle}
\usepackage{float}
\usepackage{color}

\usepackage{graphicx}
\usepackage{amssymb}
\journal{Physics Letters A}

\begin{document}

\begin{frontmatter}



\title{Nonlinearity managed vector solitons}


\author{F. Kh. Abdullaev$^{a,b}$}
\author{J. S. Yuldashev$^{a,b}$}
\author{M. \"{O}gren$^{c,d}$}
\affiliation{organization={Physical-Technical Institute, Uzbek Academy of Sciences},
            city={Tashkent},
            postcode={100084}, 
            country={Uzbekistan}}

\affiliation{organization={Theoretical Physics Department, National University of Uzbekistan},
            city={Tashkent},
            postcode={100084}, 
            country={Uzbekistan}}

\affiliation{organization={School of Science and Technology, Örebro University},
            city={Örebro},
            postcode={70182}, 
						country={Sweden}}



\affiliation{organization={HMU Research Center, Institute of Emerging Technologies},
            city={Heraklion},
            postcode={GR-71004}, 
            country={Greece}}


\begin{abstract}
The evolution of vector solitons under nonlinearity management is studied.
The averaged over strong and rapid modulations in time of the inter-species interactions vector Gross-Pitaevskii equation (GPE) is derived. 
The averaging gives the appearance of the effective nonlinear quantum pressure depending on the population of the other component. Using this system of equations, the existence and stability of the vector solitons under the action of the strong nonlinearity management (NM) is investigated. Using a variational approach the parameters of NM vector solitons are found. 
The numerical simulations of the full time-dependent coupled GPE confirms the theoretical predictions.
\end{abstract}



\end{frontmatter}







\section{Introduction}
\label{sec_1}
The properties of an atomic Bose-Einstein
condensate (BEC) under the modulation in time of a scattering length have attracted large attention the last years~\cite{Malomed1,Malomed2}.
The existence of two-dimensional bright solitons, in one and two component attractive condensates~\cite{Abd1,SU,Mont1,Mont2}, in media with competing nonlinearities~\cite{MalomedOS,Abd_pre,Abd_opt}, long lived Bloch oscillations of gap solitons~\cite{Sal1},
Faraday waves~\cite{Nicolin}, and other interesting phenomena, has been predicted and observed.
Recently it has been shown, that the strong and rapid periodic modulation in time of the scattering length (so called Feshbach resonance management) in a BEC can lead to the existence of stable compactons~\cite{Sal2}.
We are here interested in investigating if such structures also can exists in two-component BECs with equal inter-species interactions.
In distinction from the scalar case, effects of Feshbach resonance management can then be on inter- and intra-species scattering lengths,
leading to richer varieties of Faraday waves and localized states in such systems.

Mathematically the problem is reduced to the investigation of modulational instability and localized states in two coupled nonlinear Schr\"odinger equations (NLSE) with nonlinearity management of the self- and cross-phase modulation terms.

This paper is devoted to the investigation of these problems. First we consider the modulational instability of nonlinear plane waves under periodic modulations in time of the nonlinearity. 
The analysis will be performed for the case of rapid and strong modulations. We derive the averaged vector-GPE, which is called vector-NLSE in optics, with effectively nonlinear dispersion terms.
  Based on this averaged equation, we analyse the conditions of the existence of bright solitonic states and study the properties of these solutions. 
	By the numerical simulations of the original vector-GPE with nonlinearity management, we verify the
analytical predictions, including the averaged equation approach, and analyze the results further than the analytical predictions.

\section{Model}
\label{sec_2}
Two-component BECs are here described by the system of coupled Gross-Pitaevskii equations:
\begin{eqnarray}
i\hbar\psi_{1,\tilde{t}} =-\frac{\hbar^2}{2m_1}\psi_{1,\tilde{x}\tilde{x}} + (\tilde{g}_{11} |\psi_1|^2 + \tilde{g}_{12}|\psi_2|^2)\psi_1 \nonumber \\
i\hbar\psi_{2,\tilde{t}} =-\frac{\hbar^2}{2m_2}\psi_{2,\tilde{x}\tilde{x}} + (\tilde{g}_{22} |\psi_2|^2 + \tilde{g}_{12}|\psi_1|^2)\psi_2.
\end{eqnarray}
Here $\tilde{g}_{ij}=2\hbar\omega_{\perp}a_{ij}$, where $\omega_{\perp}$ is the transverse frequency of the trap, and $a_{ij}$ are the inter- and intra-species atomic scattering lengths. 
$\psi_{1,2}$ represent the wavefunctions of the individual components of a BEC.
Considering the case of equal masses $m = m_1=m_2$, for the atoms of both components, and introducing the dimensionless variables according to:
$$
x = \tilde{x}/l_{\perp}, \,\, t= \tilde{t}\omega_{\perp}, \,\, u = \sqrt{2l_{\perp}}\psi_1, \,\, v = \sqrt{2l_{\perp}}\psi_2, \,\, g_{ij} = \frac{\tilde{g}_{ij}}{2\hbar \omega_{\perp}l_{\perp}}, \,\, l_{\perp}=\sqrt{\frac{\hbar}{m \omega_{\perp}}}, 
$$
we obtain the system of a coupled GPE, without an external potential, that describe the vector solitons:
\begin{eqnarray}\label{eq:original_model}
i\frac{\partial u}{\partial t}+ \frac{1}{2}\frac{\partial^2u}{\partial x^2} - (g_{11}|u|^2 +g_{12}|v|^2)u&=&0  \nonumber \\
i\frac{\partial v}{\partial t}+ \frac{1}{2}\frac{\partial^2v}{\partial x^2} - (g_{22}|v|^2+ g_{12}| u|^2)v&=&0.
\end{eqnarray}
With the coefficients $g_{12}=\gamma$, we have the corresponding Hamiltonian:
\begin{equation}\label{eq:Hamiltonian_original}
H=\int_\mathbb{R} \left\lbrace \frac{1}{2}|u_x|^2 + \frac{1}{2}|v_x|^2 +\frac{1}{2} g_{11}|u|^4+\frac{1}{2} g_{22}|v|^4 + \gamma|u|^2|v|^2 \right\rbrace dx.
\end{equation}
In the field of optics, when studying spatial solitons, $t$ represents the propagation coordinate, while the $u$ and $v$ variables denote two mutually incoherent beams.
We consider the case of rapidly changing time dependent coefficients, that is $g_{12}(t)= \gamma(t)=\gamma_0+\gamma_1(t)=\gamma_0 + \gamma_1\cos(\omega t)$, where $\gamma_1,\omega \sim 1/\epsilon, \epsilon \ll 1$.

To obtain averaged equations, we use the following transformations~\cite{Sal2,ZharPel,Kevrekidis1,Kevrekidis2}, which removes the rapidly and strongly varying terms from the GPE, and allows to obtain the averaged over rapid modulations system of equations:
\begin{equation}\label{eq:transformations}
u = \bar{u}e^{-i\Gamma(t)|\bar{v}|^2}, \,\,\,\,\,\, v = \bar{v}e^{-i\Gamma(t)|\bar{u}|^2},
\end{equation}
where $\Gamma_t(t)=\gamma_1(t)$, i.e., $\Gamma(t)=\frac{\gamma_1}{\omega}\sin(\omega t)$. By inserting the new field transformations (\ref{eq:transformations}) into Eqs.~(\ref{eq:original_model}), we get the following equations
\begin{eqnarray}\label{eq:trans_1}
&iu_t+\Gamma(t)u|v|^2_t+\frac{1}{2}u_{xx}-i\Gamma(t)u_x|v|^2_x-i\frac{1}{2}\Gamma(t)u|v|^2_{xx}-\frac{1}{2}\Gamma^2(t)\left(|v|^2_x \right)^2u & \nonumber \\ &- \left(g_{11}|u|^2+\gamma_0|v|^2 \right)u=0 & \nonumber \\
& iv_t+\Gamma(t)v|u|^2_t+\frac{1}{2}v_{xx}-i\Gamma(t)v_x|u|^2_x-i\frac{1}{2}\Gamma(t)v|u|^2_{xx}-\frac{1}{2}\Gamma^2(t)\left(|u|^2_x \right)^2v & \nonumber \\ &- \left(g_{22}|v|^2+\gamma_0|u|^2 \right)v=0.
\end{eqnarray}
Here and hereafter we omit the bar signs for simplicity. 
It follows from Eqs.~(\ref{eq:trans_1}) that
\begin{eqnarray}\label{eq:trans_2}
|u|^2_t&=&\frac{i}{2}(u^\ast u_{xx}-u u^\ast_{xx}) + \Gamma(t)|u|^2_x|v|^2_x + \Gamma(t)|u|^2|v|^2_{xx} \nonumber \\
|v|^2_t&=&\frac{i}{2}(v^\ast v_{xx}-v v^\ast_{xx}) + \Gamma(t)|u|^2_x|v|^2_x + \Gamma(t)|v|^2|u|^2_{xx}.
\end{eqnarray}
After inserting Eqs.~(\ref{eq:trans_2}) into the transformed equations~(\ref{eq:trans_1}), and averaging over the period of rapid oscillations $\Lambda=2\pi/\omega$, we derive the following set of coupled averaged equations:
\begin{eqnarray}\label{eq:averaged_model}
&iu_t+\sigma^2\left[|u|^2_x|v|^2_x + |v|^2|u|^2_{xx} - \frac{1}{2}\left(|v|^2_x \right)^2 \right]u+\frac{1}{2}u_{xx}& \nonumber\\
&- \left(g_{11}|u|^2+\gamma_0|v|^2 \right)u=0&  \nonumber\\
&iv_t+\sigma^2\left[|u|^2_x|v|^2_x + |u|^2|v|^2_{xx} - \frac{1}{2}\left(|u|^2_x \right)^2 \right]v+\frac{1}{2}v_{xx}& \nonumber\\
&- \left(g_{22}|v|^2+\gamma_0|u|^2 \right)v=0,&
\end{eqnarray}
where  $\sigma^2=\left\langle \Gamma(t)^2 \right\rangle = \gamma_1^2/(2\omega^2)$. 
Hence, the standard Hamiltonian (\ref{eq:Hamiltonian_original}) can now be written in the following averaged form:
%
%
\begin{eqnarray}\label{eq:Hamiltonian_averaged}
\bar{H}=\int_\mathbb{R} \lbrace \frac{1}{2}|u_x|^2 + \frac{1}{2}|v_x|^2 &+& \frac{1}{2}\sigma^2 \left(|u|^2\left(|v|^2_x \right)^2 +|v|^2\left(|u|^2_x \right)^2 \right)  +\frac{1}{2} g_{11}|u|^4   \nonumber\\
&+&\frac{1}{2}g_{22}|v|^4 + \gamma_0|u|^2|v|^2 \rbrace dx.
\end{eqnarray}
The averaged system (\ref{eq:averaged_model}) shows that the effective nonlinear quantum pressure appears in addition to the original linear quantum pressure. The magnitude of this correction depends on the population of the other component. 

The appearance of the nonlinear dispersive terms will lead to new effects in the dynamics of the matter waves. 
In particular to the existence of nonlinearity managed vector solitons. 
Standard vector solitons, i.e. in one component, exists due to the balance between the second-order linear dispersion (the quantum pressure) and the mean-field cubic nonlinearity. 
Now the balance between linear dispersion, nonlinear dispersion and the mean-field nonlinearity can give rise to the existence of NM vector solitons. 

In the next sections we will consider the modulational instability in the NM system, which is important for defining the region of parameters where solitons can be generated.  The existence and stability of NM vector solitons will also be considered.

\begin{figure}[H]
\begin{center}
\includegraphics[scale=0.21]{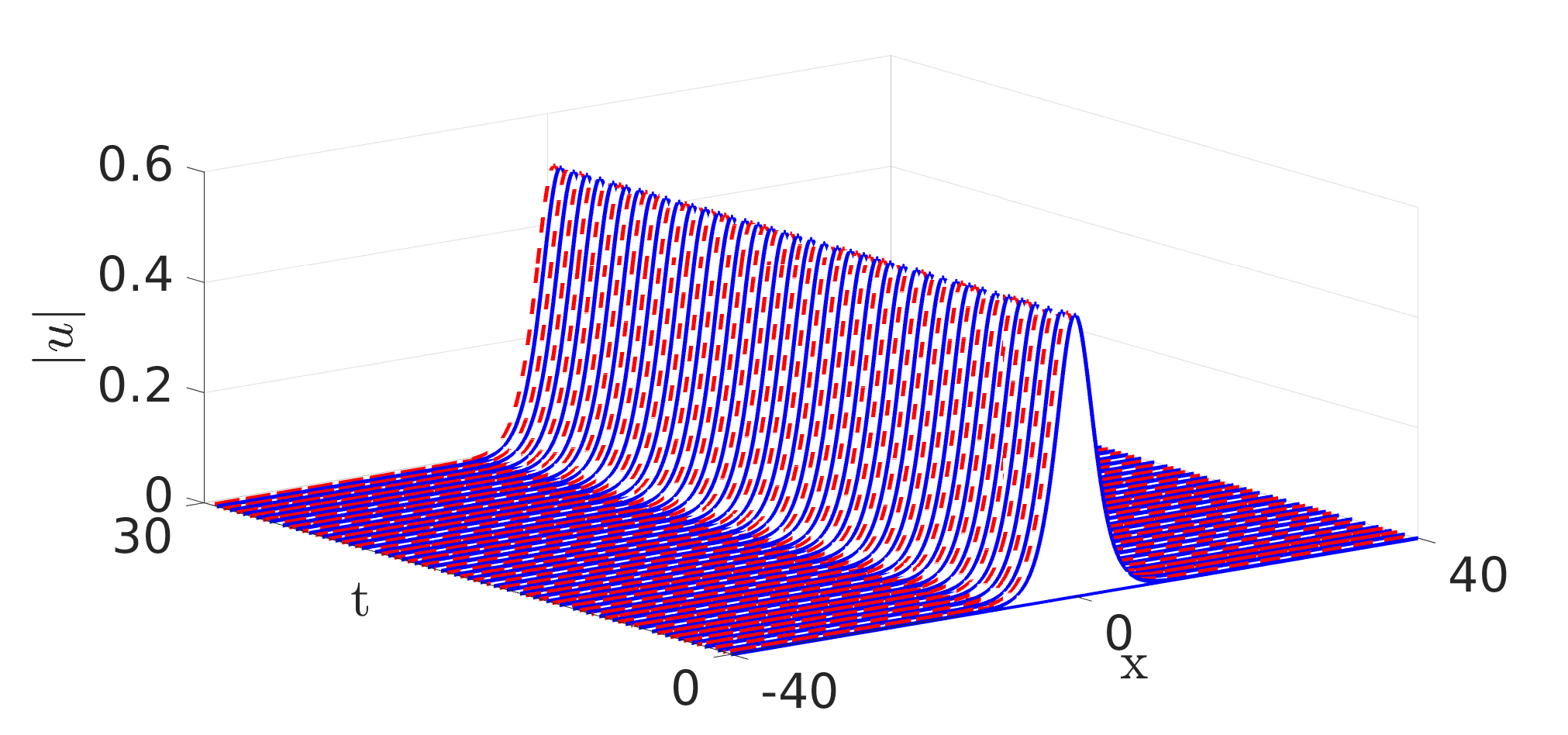}~\hspace{-4mm}~\includegraphics[scale=0.21]{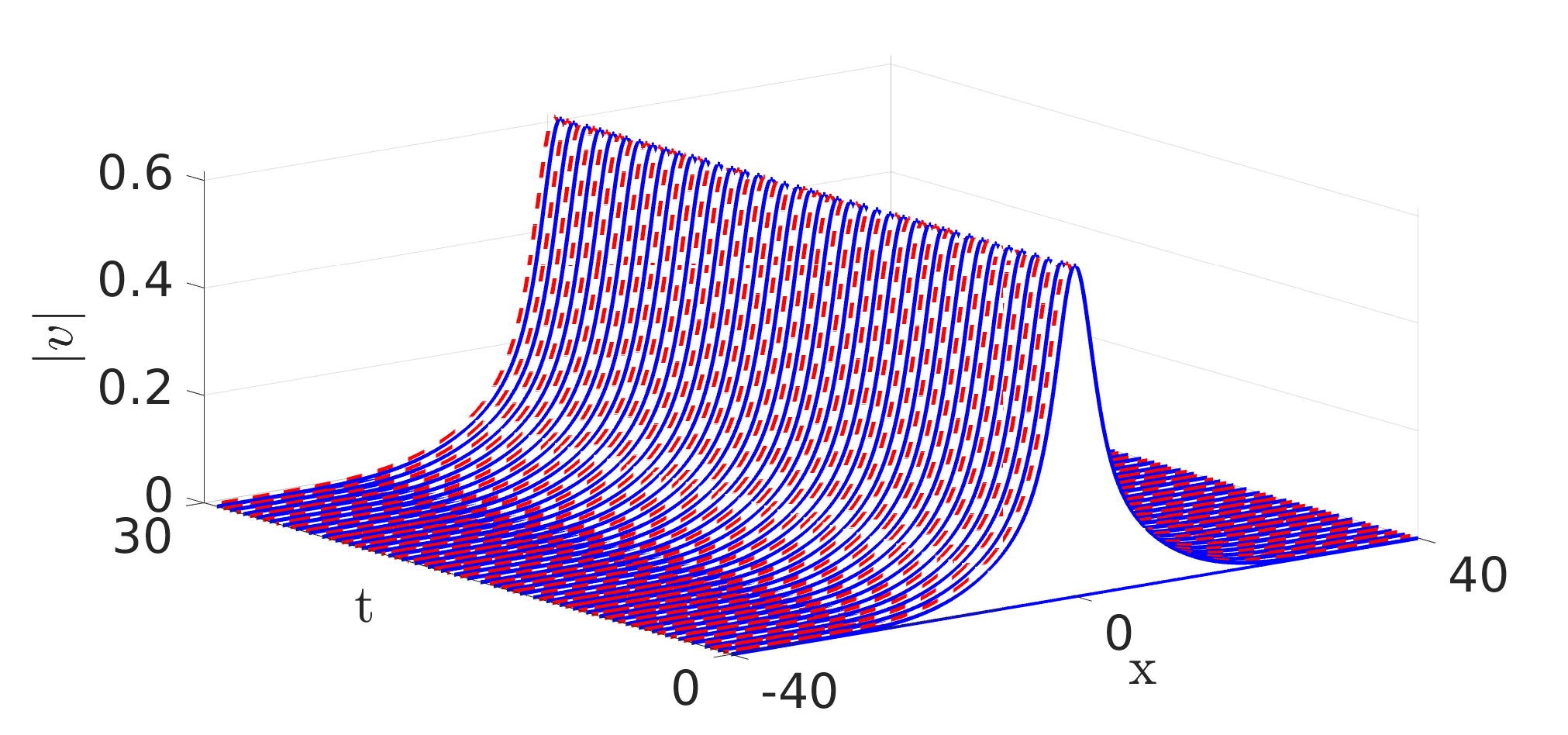}
\includegraphics[scale=0.31]{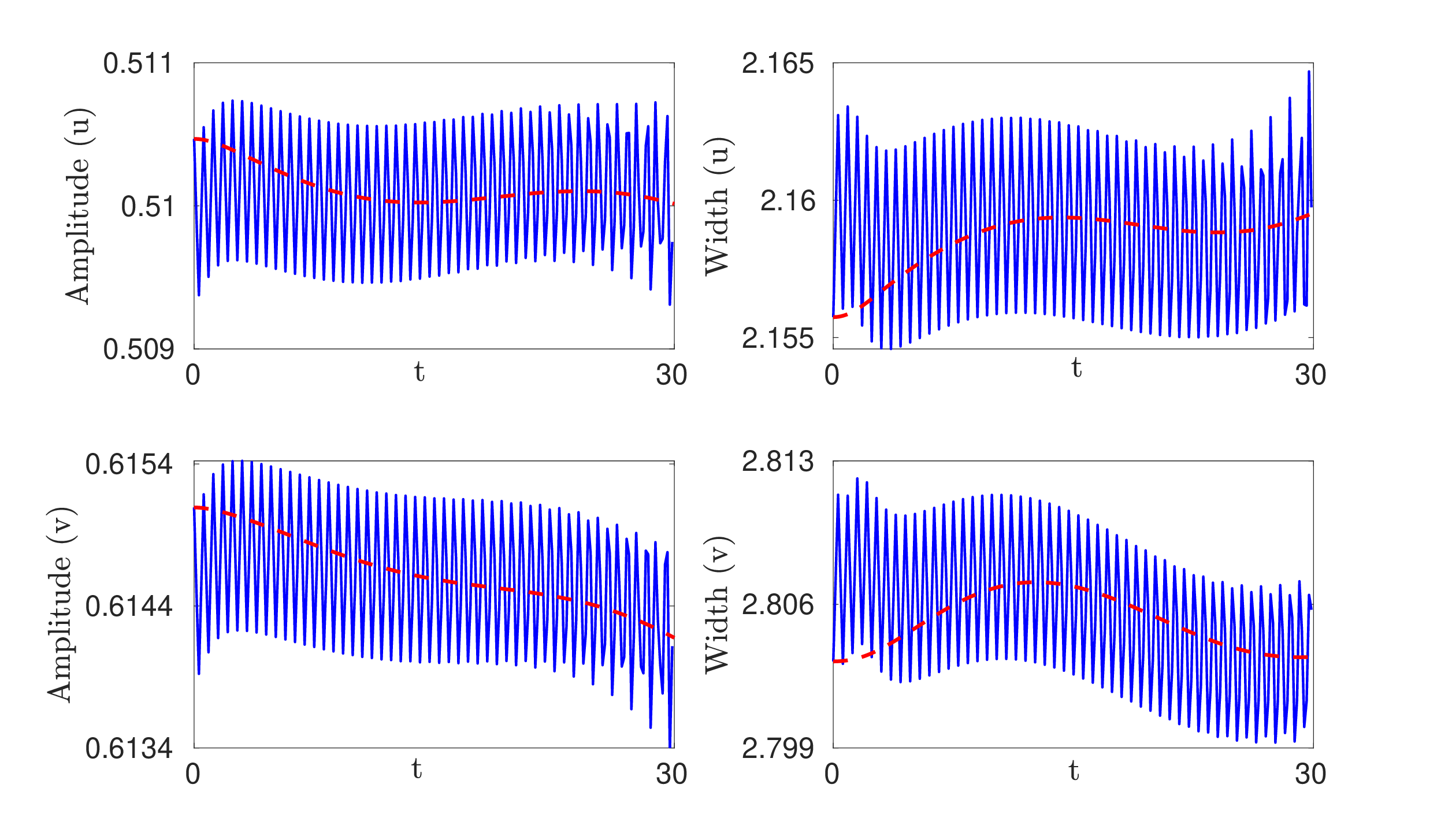}
\caption{Comparison of the averaged Eqs.~(\ref{eq:averaged_model}) and the original model (\ref{eq:original_model}). The top frames display the temporal evolution of soliton components ($u$ and $v$), with the averaged model shown as blue solid curves and the original model as red dashed curves. The middle frames show the amplitude and width of the $u$ component, while the bottom frames depict the amplitude and width of the $v$ component. In all frames, the dashed red curves correspond to the averaged model, while the solid blue curves represent the original model. Norms of the corresponding components are: $N_u = 1$ and $N_v=2$. Parameters used are: $g_{11}=1.4$, $g_{22}=1.1$, $\gamma_0=-2$, $\gamma_1=15$, and $\omega=10 \pi$.}
\label{fig:1}
\end{center}
\end{figure}

To assess the accuracy of the averaged model, we conducted numerical simulations of both the original system~(\ref{eq:original_model}) and the averaged equations~(\ref{eq:averaged_model}). 
The simulation utilized the same initial condition that was obtained by numerically solving for the stationary states of Eqs.~(\ref{eq:averaged_model}). An example of the comparative analysis of the stationary solutions' evolution is depicted in Fig.~\ref{fig:1}. There we used the parameters $\omega = 10\pi$, $\gamma_0=-2$, and $\gamma_1=15$, for which we illustrate the comparison. We have also successfully verified the consistency between the results obtained from the original and averaged models for other sets of parameters with moderate values of $\gamma_1$. 
In all the following graphs, the numerical widths are calculated using the relation $\textit{width} = \textit{fwhm}/\sqrt{8\ln 2}$, which represents the relationship between the actual \textit{width} of a Gaussian beam and the full width at half maximum, denoted as \textit{fwhm}.

\section{Modulational Instability}
\label{sec_3}
In this section, we consider modulational instability of matter waves in two-component BEC under nonlinearity management. Analytical considerations will be performed using the averaged system and then compared to numerical simulations of the full time dependent model. To do so, we first consider constant amplitude solutions of Eqs.~(\ref{eq:averaged_model}) of the form
\begin{equation}\label{eq:cw_solution}
u=Ae^{i(q_ux+\omega_ut)}, \,\,\,\,\, v=Be^{i(q_vx+\omega_vt)},
\end{equation}
where $A$ and $B$ are constants, and $q_u$, $q_v$, $\omega_u$, and $\omega_v$ satisfy the dispersion relations
\begin{equation}\label{eq:dispersion_relations}
\omega_u=-\frac{1}{2}q^2_u-(g_{11} A^2 + \gamma_0 B^2), \,\,\,\, \omega_v=-\frac{1}{2}q^2_v-(g_{22} B^2 + \gamma_0 A^2).
\end{equation}
We now study the stability of the constant amplitude solutions against small perturbations. Hence, we look for solutions in the form 
\begin{equation}\label{eq:perturbations}
u=\left[A+\delta A(x,t)\right]e^{i(q_ux+\omega_ut)}, \,\,\,\,\, v=\left[B+\delta B(x,t)\right]e^{i(q_vx+\omega_vt)},
\end{equation}
where $\delta A(x,t)$ and $\delta B(x,t)$ are complex functions that represents small perturbations. 
Substituting Eqs.~(\ref{eq:perturbations}) into the averaged equations~(\ref{eq:averaged_model}), taking into account the dispersion relations~(\ref{eq:dispersion_relations}) and linearizing the resulting equations, we obtain the following two coupled equations: 
\begin{eqnarray}\label{eq:perturbations_2}
& i\frac{\partial \delta A}{\partial t}+\sigma^2 B^2A^2\left(\frac{\partial^2 \delta A^\ast}{\partial x^2} + \frac{\partial^2 \delta A}{\partial x^2} \right) + \frac{1}{2}\frac{\partial^2 \delta A}{\partial x^2} & \nonumber \\ & -\left[ g_{11} \left(\delta A^\ast + \delta A\right) A^2 +\gamma_0 \left(\delta B^\ast + \delta B\right)BA \right] =0   \nonumber \\ 
& i\frac{\partial \delta B}{\partial t}+\sigma^2 B^2A^2\left(\frac{\partial^2 \delta B^\ast}{\partial x^2} + \frac{\partial^2 \delta B}{\partial x^2} \right) + \frac{1}{2}\frac{\partial^2 \delta B}{\partial x^2} & \nonumber \\ & - \left[ g_{22} \left(\delta B^\ast + \delta B\right) B^2 +\gamma_0 \left(\delta A^\ast + \delta A\right)BA \right]=0.
\end{eqnarray}
Decomposing the perturbations into real and imaginary parts 
and looking for solutions of these unknown functions in the form of plane waves $\exp(i\Omega t + ikx)$, we obtain the following dispersion relation:
\begin{equation}\label{eq:dispersion_relation_2}
\Omega^2=\frac{1}{2}\left[f_1 +f_2 \pm \sqrt{(f_1-f_2)^2 + 4C^2} \right],
\end{equation}  
where $f_1=k^2\left[k^2(1+4\sigma^2A^2B^2)/4 + g_{11}A^2 \right]$, $f_2=k^2\left[k^2(1+4\sigma^2A^2B^2)/4 + g_{22}B^2 \right]$, and $C^2=\gamma^2_0k^4A^2B^2$.
The modulational instability (MI) occurs when $\Omega^2 < 0$ for any wavenumbers $k$. 
For small $k$ we then get the following condition from Eq.~(\ref{eq:dispersion_relation_2}):
\begin{equation}
\gamma_0^2 > g_{11}g_{22}.
\end{equation}\label{eq:MI} 
The growth rate $G$, also called the gain of the modulational instability, is related to the imaginary part of $\Omega$, and is given by
\begin{equation}\label{eq:gain}
G(k)=\left(\frac{1}{2}\left[\sqrt{(f_1-f_2)^2 + 4C^2} - f_1 -f_2 \right]\right)^{\frac{1}{2}}.
\end{equation}
When considering small values of $k$, the growth rate (\ref{eq:gain}) exhibits an absolute maximum, which corresponds to the maximum gain. By applying Fermat's theorem $d G(k)/dk=0$, we can determine the maximum gain ($G_{\max}$) and its corresponding wavenumber ($k_0$):
\begin{equation}\label{eq:k_max}
k_0 = \left(\sqrt{\left( g_{11}A^2-g_{22}B^2\right)^2 + 4\gamma_0^2A^2B^2} - g_{11}A^2-g_{22}B^2 \right)^{\frac{1}{2}}\cdot \frac{1}{\sqrt{1+4\sigma^2A^2B^2}},
\end{equation}
\begin{equation}\label{eq:G_max}
G_{\max} = \left(\sqrt{\left( g_{11}A^2-g_{22}B^2\right)^2 + 4\gamma_0^2A^2B^2} - g_{11}A^2-g_{22}B^2 \right)^{\frac{1}{2}} \frac{|k_0|}{2}.
\end{equation}
From Eqs.~(\ref{eq:k_max}) and~(\ref{eq:G_max}), it becomes evident that as the parameter $\sigma^2$ increase, which is related to the modulation, both the maximum gain $G_{\max}$ and the corresponding wavenumber $k_0$ decrease in accordance with the factor $1/\sqrt{1+4 \sigma^2 A^2B^2}$. 

\begin{figure}[H]
\begin{center}
\includegraphics[scale=0.37]{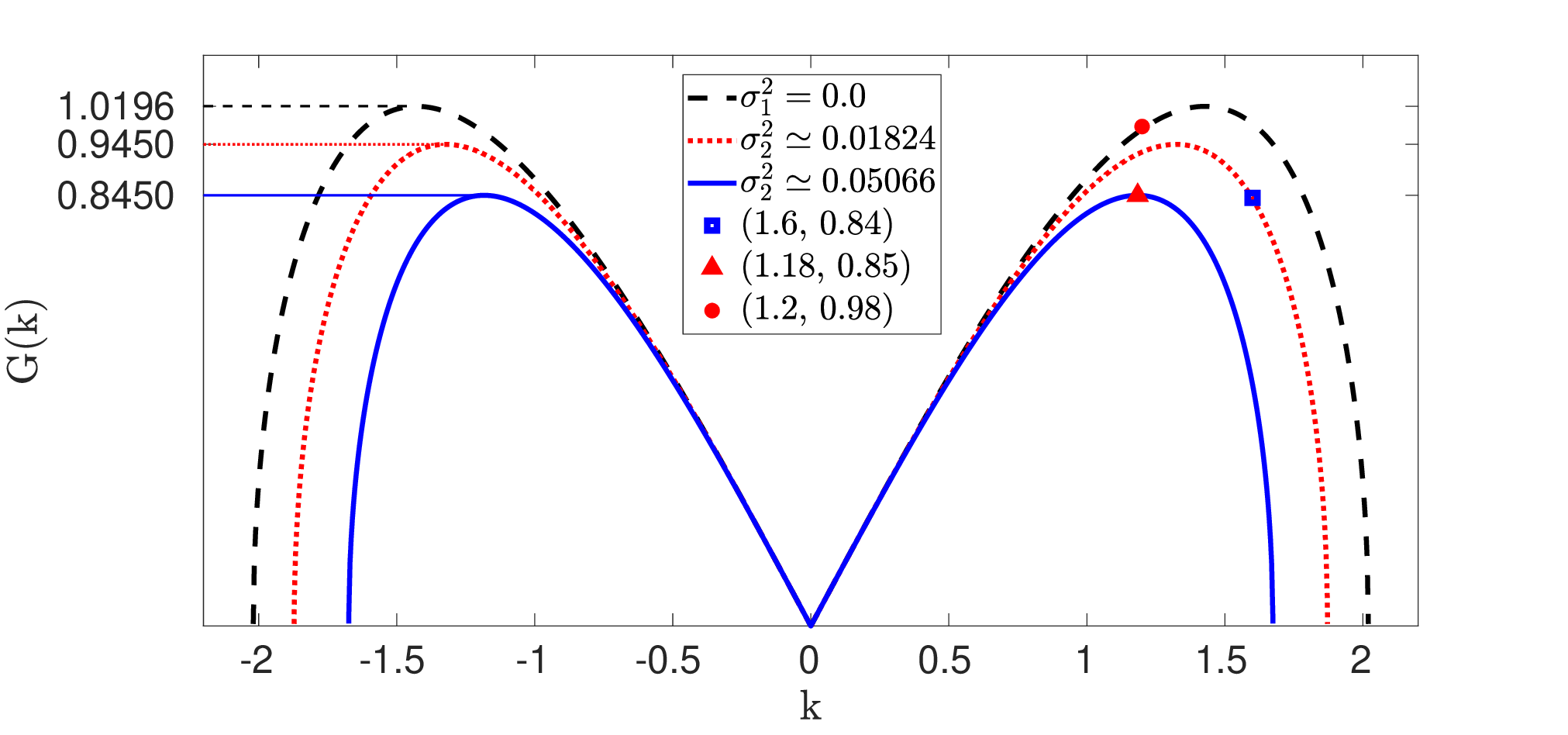} 
\caption{The gain spectrum of Eq.~(\ref{eq:gain}) for modulation instability, for three different values of $\gamma_1=0; 6; 10$ ($\sigma^2 \simeq 0.0$; $0.01824$; $0.05066$). Other parameter values are: $g_{11}=1.2$, $g_{22}=1.4$, $\gamma_0=-2$, and $\omega=10 \pi$.}
\label{fig:5}
\end{center}
\end{figure}

\begin{figure}[H]
\begin{center}
\includegraphics[scale=0.21]{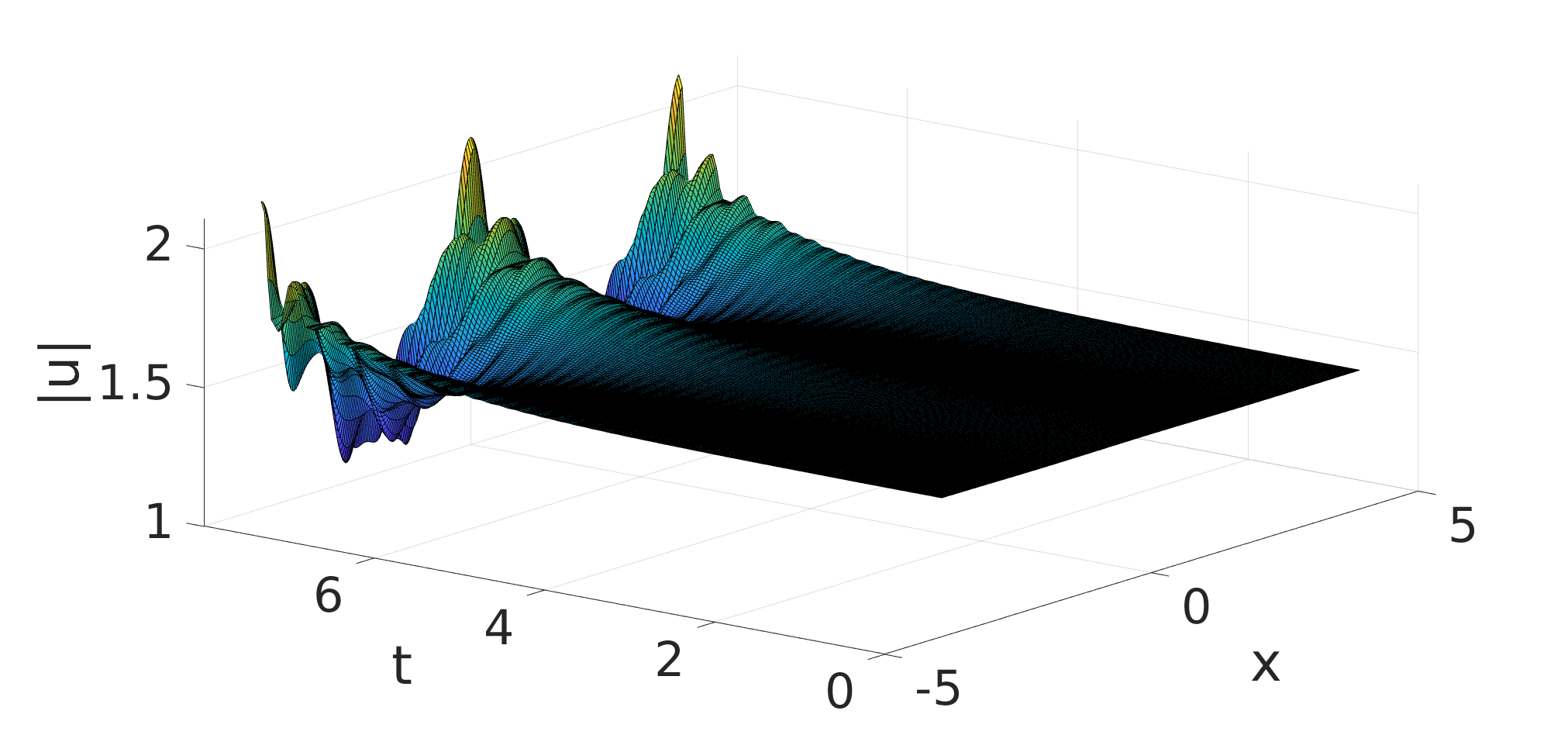}~\includegraphics[scale=0.21]{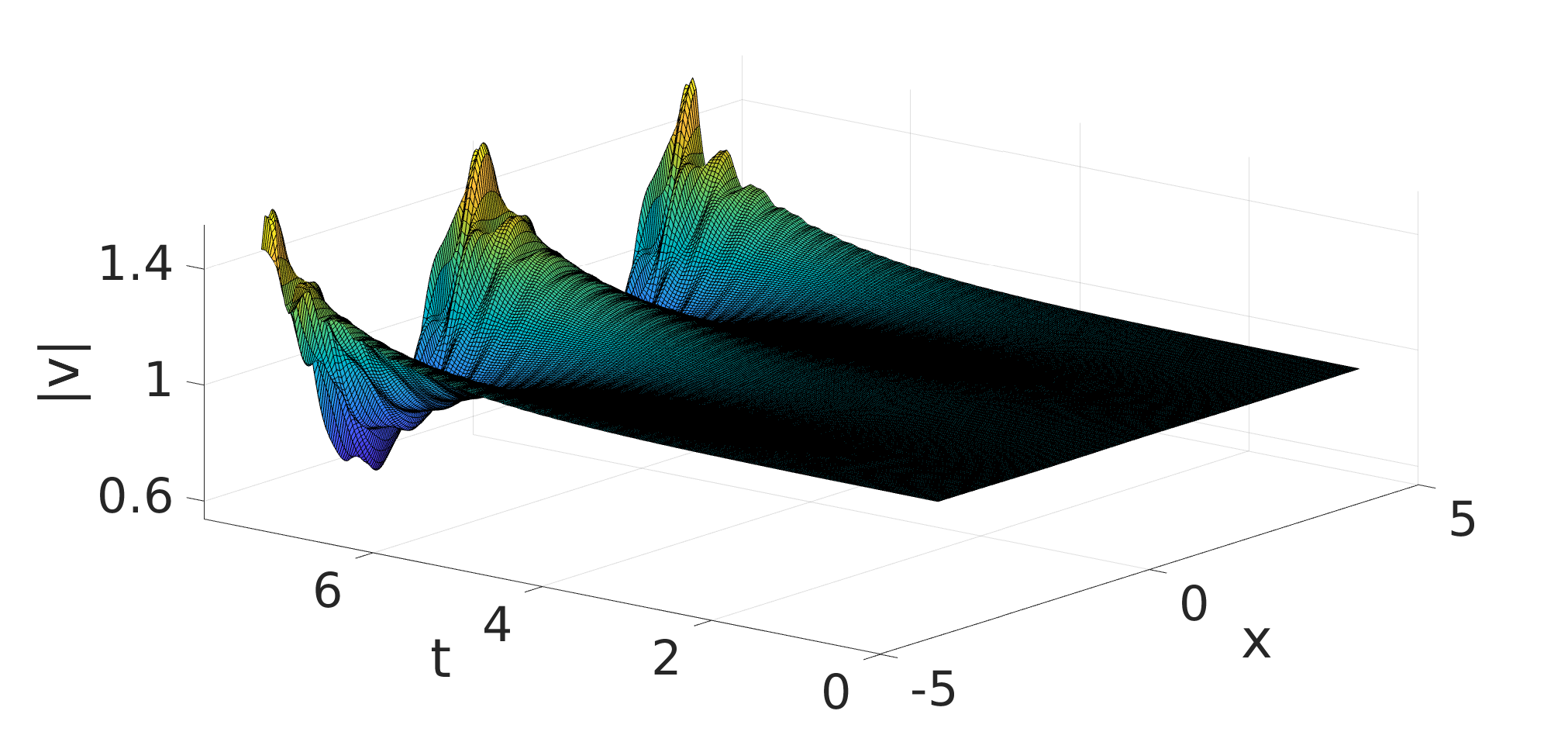}
\includegraphics[scale=0.21]{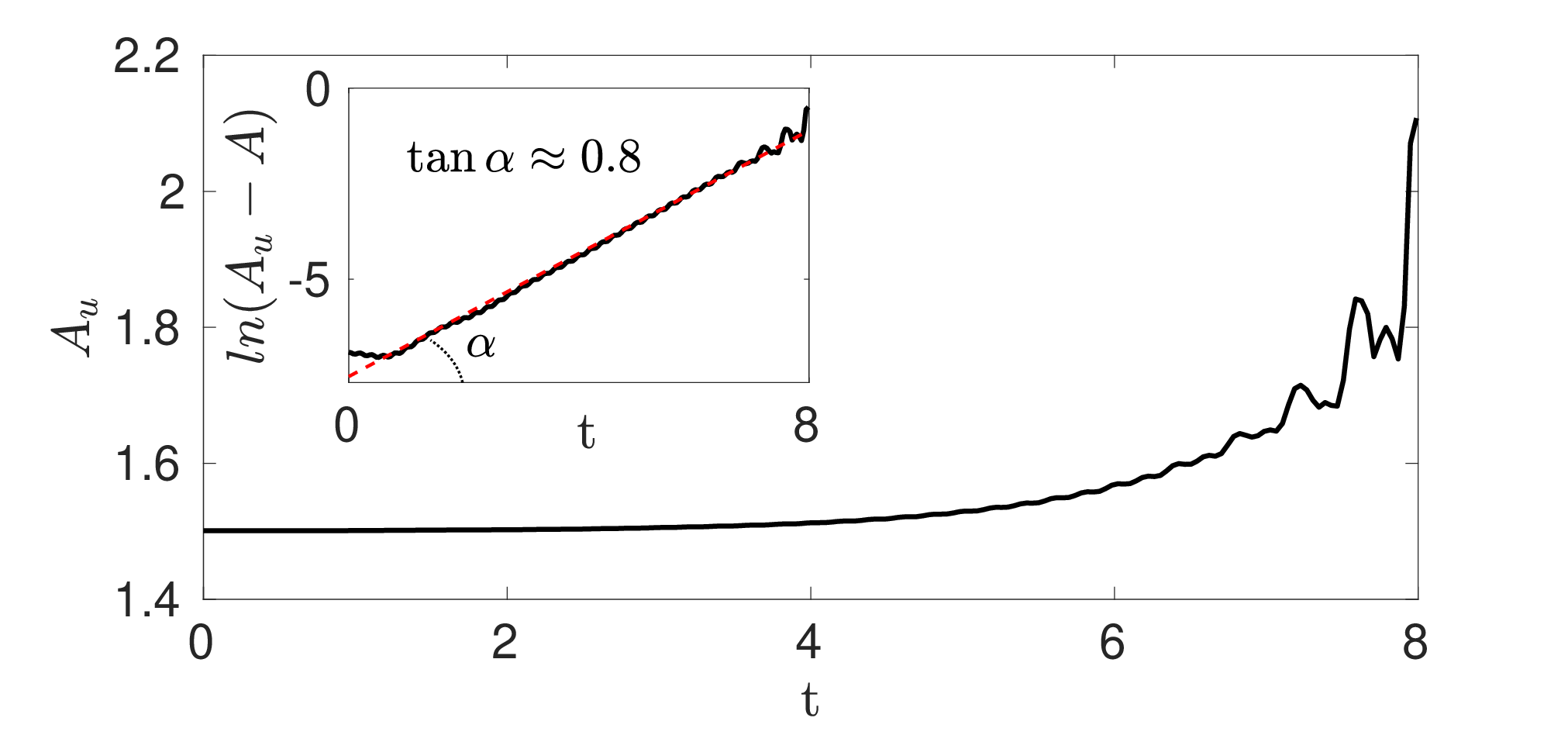}~\includegraphics[scale=0.21]{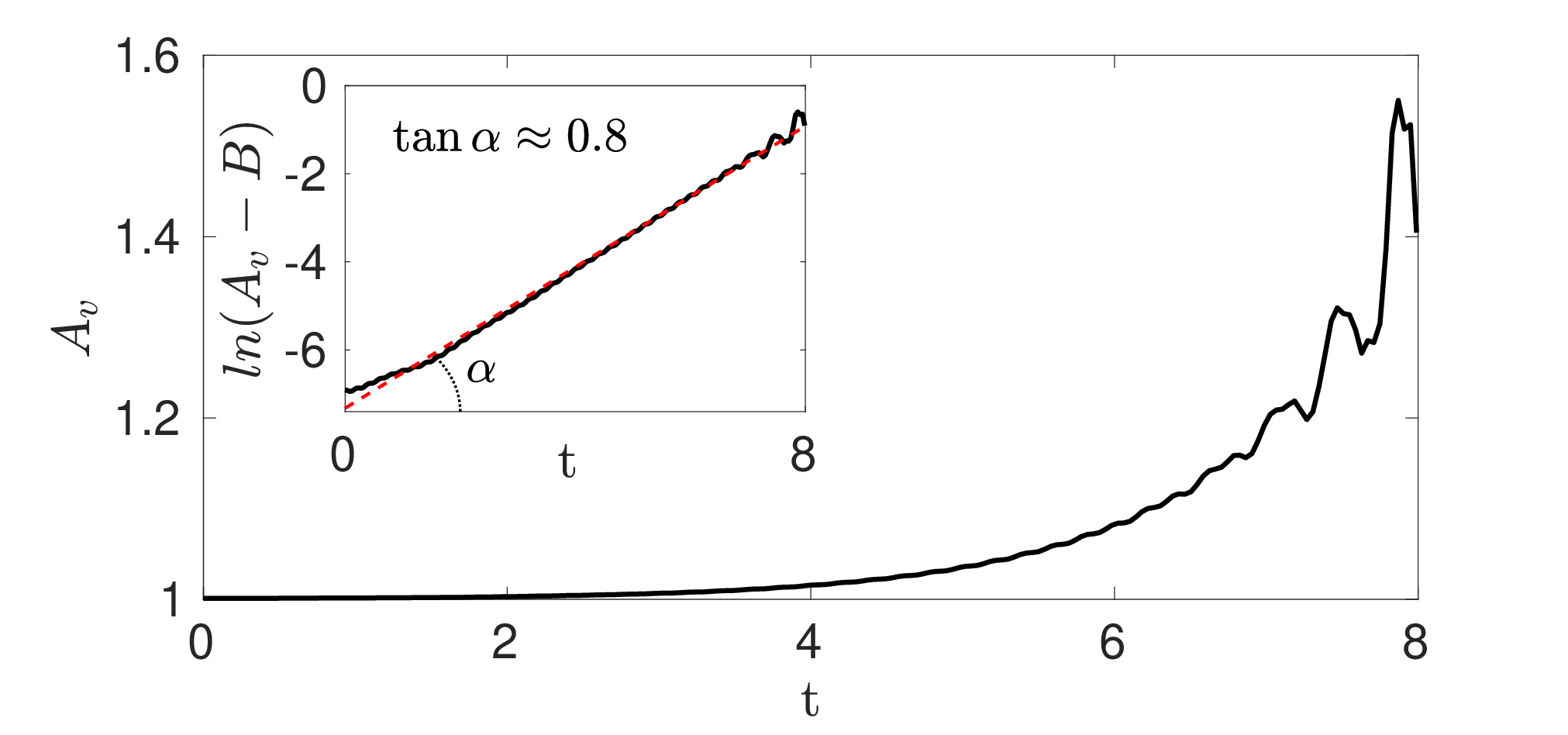}
\caption{The evolution of the plane waves (\textit{upper figures}), and their maximum amplitudes $A_u=\max(|u|)$ and $A_v = \max(|v|)$ (\textit{bottom figures}), in the modulationally unstable region with the initial perturbation $u_0=A+0.001\cos(kx)$ and $v_0=B+0.001\cos(kx)$. Where $k=1.6$ corresponds to the (blue) square depicted in Fig.~\ref{fig:5}. 
The inset shows the exponential growth rates of the amplitudes ($\approx 0.8$ for both $A_u$ and $A_v$) found from solving Eq.~(\ref{eq:original_model}) numerically, which is very close to the analytical prediction ($\approx 0.84$, see (blue) square on the $\gamma_1=6$ curve in Fig.~\ref{fig:5}). 
Parameters used are: $g_{11}=1.2$, $g_{22}=1.4$, $\gamma_0=-2$, $\gamma_1=6$, and $\omega=10 \pi$.}
\label{fig:6}
\end{center}
\end{figure}

The curves plotted in Fig.~\ref{fig:5} illustrates the relationship between the wave number of the perturbation $k$, and the gain, as defined by Eq.~(\ref{eq:gain}). 
We used the values $A=1.5$ and $B=1$, with three different modulational parameter values ($\sigma^2$). 
Studying Fig.~\ref{fig:5}, we can verify that the relationship between modulation strength and maximum gain adheres to the ratio $1.0196:0.9450:0.8450 = \sqrt{1+9\sigma_1^2}:\sqrt{1+9\sigma_2^2}:\sqrt{1+9\sigma_3^2}$, i.e., as the last factor in~(\ref{eq:k_max}).
We have performed numerical simulations of the original model~(\ref{eq:original_model}) using the same parameters as in Fig.~\ref{fig:5}. 
We compare the analytically found growth rates, which are marked at different points with different styles (see the legend of Fig.~\ref{fig:5}), to growth rates obtained through numerical solutions of the original model~(\ref{eq:original_model}). 
Specifically, we use perturbed plane waves with the frequencies ($k$) corresponding to the marked points in Fig.~\ref{fig:5} as initial conditions $u_0$ and $v_0$. 
In Fig.~\ref{fig:6}, it is shown that the analytical predictions corresponding to the (blue) square in Fig.~\ref{fig:5} is confirmed by full simulations of the original model~(\ref{eq:original_model}). 
For the rest of the marked points in Fig.~\ref{fig:5}, numerically calculated growth rates, corresponding to spatial frequencies $k=1.18$ (red triangle), and $1.2$ (red dot) are $G=0.85$, and $0.98$, respectively. 

\section{Variational approach}
\label{sec_4}
\begin{figure}[H]
\begin{center}
\includegraphics[scale=0.32]{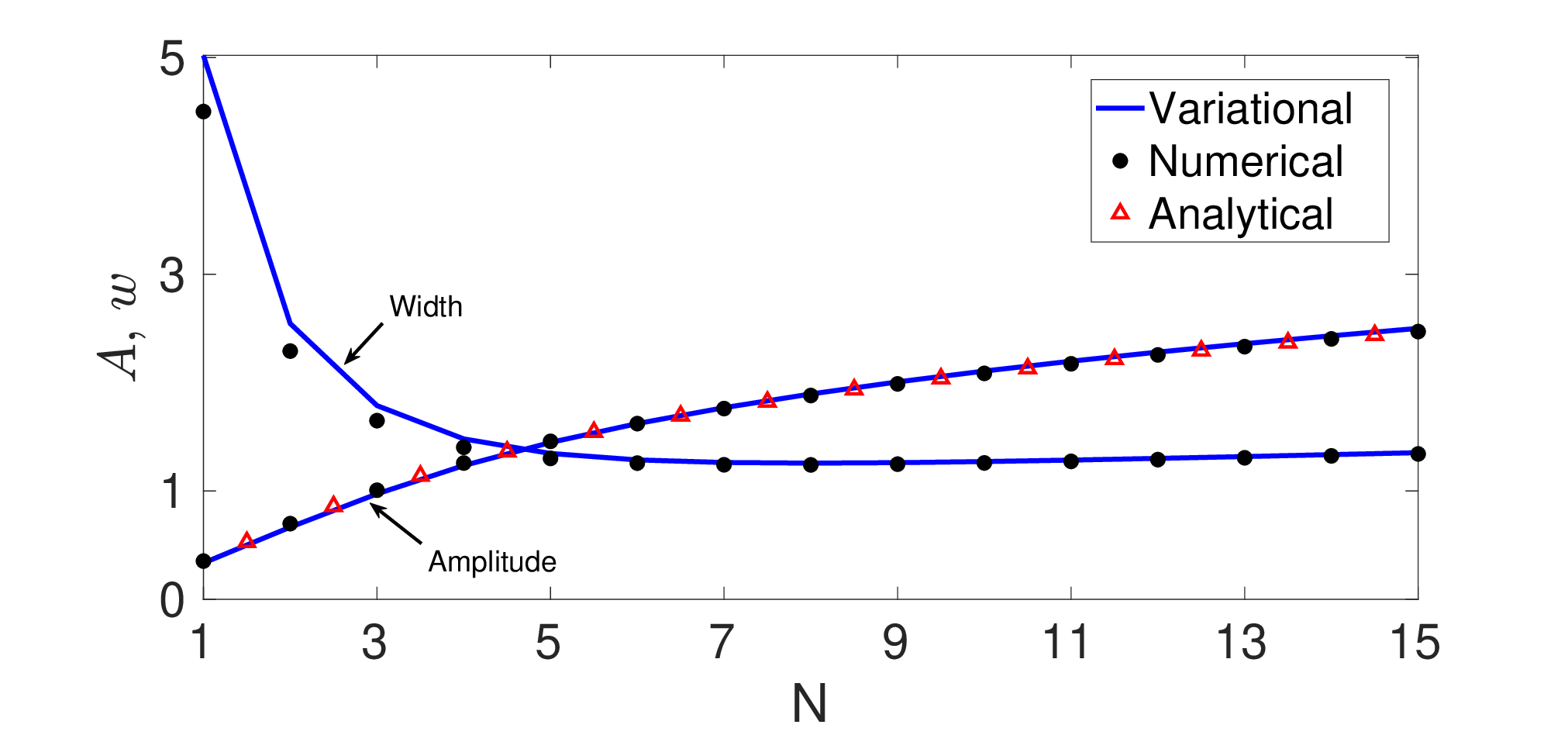} 
\caption{Widths ($w$) and amplitudes ($A$) of stationary solutions versus the norm ($N$). The (black) dots correspond to the values obtained through numerical solutions of Eq.~(\ref{eq:averaged_model}), the (red) triangles correspond to the amplitudes analytically obtained by Eq.~(\ref{eq:amplitude_analytical}), and the (blue) solid curves correspond to the values of $w$ obtained by Eq.~(\ref{eq:roots}), and $A=\sqrt{N/(\sqrt{\pi}w)}$. 
Parameters used are: $g=1.5$, $\gamma_0=-2$, $\gamma_1=10$, and $\omega=10 \pi$.}
\label{fig:11}
\end{center}
\end{figure}

Let us consider the existence and stability of solitons in the nonlinearity managed vector-GPE. 
The existence follows from the results of the existence of the modulational instability, considered in the Section~\ref{sec_3}.
It is useful to employ the variational approach 
 to Eqs.~(\ref{eq:averaged_model}) for the description of the vector solitons~\cite{Anderson}.
Eqs.~(\ref{eq:averaged_model}) can be obtained by the following Lagrangian density
\begin{eqnarray}\label{eq:LD}
&L=\frac{i}{2}\left(u\frac{\partial u^\ast}{\partial t} - u^\ast\frac{\partial u}{\partial t} \right) + \frac{i}{2} \left|\frac{\partial u}{\partial x} \right|^2 + \frac{1}{2} g_{11} |u|^4 + \frac{i}{2}\left(v\frac{\partial v^\ast}{\partial t} - v^\ast\frac{\partial v}{\partial t} \right) & \nonumber \\ &  + \frac{i}{2} \left|\frac{\partial v}{\partial x} \right|^2 + \frac{1}{2} g_{22} |v|^4 + \gamma_0 |u|^2|v|^2 +\frac{\sigma^2}{2}\left(|u|^2\left(|v|^2_x \right)^2 +|v|^2\left(|u|^2_x \right)^2 \right).
\end{eqnarray}
We use trial Gaussian functions as ansatz functions with real time-dependent amplitudes $A_i(t)$, widths $w_i(t)$, chirps $b_i(t)$ and phases $\phi_i(t)$ ($i=1,2$)
\begin{equation}\label{eq:ansatz}
u=A_1e^{-\frac{x^2}{2w_1^2}+ib_1x^2+i\phi_1}, \,\,\,\,\, v=A_2e^{-\frac{x^2}{2w_2^2}+ib_2x^2+i\phi_2}.
\end{equation}
A set of coupled ordinary differential equations for the eight ansatz parameters ($A_i$, $w_i$, $b_i$, and $\phi_i$) can be obtained from the following variational principle: 
\begin{equation}\label{eq:variational_principle}
 \delta \int \bar{L}dx=0, 
\end{equation}
where $\bar{L}$ denotes the averaged Lagrangian density, which is obtained by integrating the result of inserting the
Gaussian functions given by (\ref{eq:ansatz}) into the Lagrangian density~(\ref{eq:LD}). By performing the integration, we obtain
\begin{eqnarray}
&\bar{L}  = \frac{\sqrt{2} N_{u}^{2} g_{11}}{4 \sqrt{\pi} w_{1}{\left(t \right)}} + \frac{N_{u} N_{v} \gamma_{0}}{\sqrt{\pi} \sqrt{w_{1}^{2}{\left(t \right)}+ w_{2}^{2}{\left(t \right)}}} + N_{u} b_{1}^{2}{\left(t \right)} w_{1}^{2}{\left(t \right)} +  \frac{N_{u} w_{1}^{2}{\left(t \right)} \frac{d}{d t} b_{1}{\left(t \right)}}{2} + N_{u} \frac{d}{d t} \phi_{1}{\left(t \right)} & \nonumber \\
& + \frac{0.25 N_{u}}{w_{1}^{2}{\left(t \right)}} + \frac{\sqrt{2} N_{v}^{2} g_{22}}{4 \sqrt{\pi} w_{2}{\left(t \right)}}+ N_{v} b_{2}^{2}{\left(t \right)} w_{2}^{2}{\left(t \right)} + \frac{N_{v} w_{2}^{2}{\left(t \right)} \frac{d}{d t} b_{2}{\left(t \right)}}{2} + N_{v} \frac{d}{d t} \phi_{2}{\left(t \right)} + \frac{0.25 N_{v}}{w_{2}^{2}{\left(t \right)}} & \nonumber \\ & + \sigma^2 \left(\frac{N_{u}^{2} N_{v} w_{2}^{2}{\left(t \right)}}{\pi \left[w_{1}^{2}{\left(t \right)} + 2 w_{2}^{2}{\left(t \right)}\right]^{1.5} w_{1}^{3}{\left(t \right)}} + \frac{N_{u} N_{v}^{2} w_{1}^{2}{\left(t \right)}}{\pi \left[2 w_{1}^{2}{\left(t \right)} + w_{2}^{2}{\left(t \right)}\right]^{1.5} w_{2}^{3}{\left(t \right)}}\right),
\end{eqnarray}
where $N_u=\sqrt{\pi}A_1^2w_1$ and $N_v=\sqrt{\pi}A_2^2w_2$. 
The equations for the Gaussian ansatz parameters $\xi_i \to  \{ A_i, w_i, b_i, \phi_i \}$ are derived from Euler-Lagrange equations $d/dt(\partial \bar{L}/\partial\dot{\xi}_i) - \partial \bar{L}/\partial\xi_i = 0$, $i = 1, 2$. 
The Euler-Lagrange equations provide conservation of norms ($N_u = \textnormal{const}$, and $N_v = \textnormal{const}$), for the phases ($\phi_i$), and they result in the following system of equations for the widths ($w_i$) and chirps ($b_i$):
\begin{eqnarray}\label{eq:width_1}
& -\frac{\sqrt{2} N_{u}^{2} g_{11}}{4 \sqrt{\pi} w_{1}^{2}{\left(t \right)}} - \frac{N_{u} N_{v} \gamma_{0} w_{1}{\left(t \right)}}{\sqrt{\pi} \left[w_{1}^{2}{\left(t \right)} + w_{2}^{2}{\left(t \right)}\right]^{\frac{3}{2}}} + 2 N_{u} b_{1}^{2}{\left(t \right)} w_{1}{\left(t \right)} + N_{u} w_{1}{\left(t \right)} \frac{d}{d t} b_{1}{\left(t \right)} - \frac{0.5 N_{u}}{w_{1}^{3}{\left(t \right)}}& \nonumber \\ 
& -\sigma^2 \left( \frac{3.0 N_{u}^{2} N_{v} w_{2}^{2}{\left(t \right)}}{\pi \left[w_{1}^{2}{\left(t \right)} + 2 w_{2}^{2}{\left(t \right)}\right]^{\frac{5}{2}} w_{1}^{2}{\left(t \right)}} + \frac{3 N_{u}^{2} N_{v} w_{2}^{2}{\left(t \right)}}{\pi \left[w_{1}^{2}{\left(t \right)} + 2 w_{2}^{2}{\left(t \right)}\right]^{\frac{3}{2}} w_{1}^{4}{\left(t \right)}} + \frac{6.0 N_{u} N_{v}^{2} w_{1}^{3}{\left(t \right)}}{\pi \left[2 w_{1}^{2}{\left(t \right)} + w_{2}^{2}{\left(t \right)}\right]^{\frac{5}{2}} w_{2}^{3}{\left(t \right)}} \right. & \nonumber \\ 
&\left. - \frac{2 N_{u} N_{v}^{2} w_{1}{\left(t \right)}}{\pi \left[2 w_{1}^{2}{\left(t \right)} + w_{2}^{2}{\left(t \right)}\right]^{\frac{3}{2}} w_{2}^{3}{\left(t \right)}}\right) =0,&
\end{eqnarray}
\begin{eqnarray}\label{eq:width_2}
&-\frac{N_{u} N_{v} \gamma_{0} w_{2}{\left(t \right)}}{\sqrt{\pi} \left[w_{1}^{2}{\left(t \right)} + w_{2}^{2}{\left(t \right)}\right]^{\frac{3}{2}}} - \frac{\sqrt{2} N_{v}^{2} g_{22}}{4 \sqrt{\pi} w_{2}^{2}{\left(t \right)}} + 2 N_{v} b_{2}^{2}{\left(t \right)} w_{2}{\left(t \right)} + N_{v} w_{2}{\left(t \right)} \frac{d}{d t} b_{2}{\left(t \right)} - \frac{0.5 N_{v}}{w_{2}^{3}{\left(t \right)}} & \nonumber \\ 
& -\sigma^2 \left(\frac{6.0 N_{u}^{2} N_{v} w_{2}^{3}{\left(t \right)}}{\pi \left[w_{1}^{2}{\left(t \right)} + 2 w_{2}^{2}{\left(t \right)}\right]^{\frac{5}{2}} w_{1}^{3}{\left(t \right)}} - \frac{2 N_{u}^{2} N_{v} w_{2}{\left(t \right)}}{\pi \left[w_{1}^{2}{\left(t \right)} + 2 w_{2}^{2}{\left(t \right)}\right]^{\frac{3}{2}} w_{1}^{3}{\left(t \right)}} + \frac{3.0 N_{u} N_{v}^{2} w_{1}^{2}{\left(t \right)}}{\pi \left[2 w_{1}^{2}{\left(t \right)} + w_{2}^{2}{\left(t \right)}\right]^{\frac{5}{2}} w_{2}^{2}{\left(t \right)}} \right. & \nonumber \\ 
& \left. + \frac{3 N_{u} N_{v}^{2} w_{1}^{2}{\left(t \right)}}{\pi \left[2 w_{1}^{2}{\left(t \right)} + w_{2}^{2}{\left(t \right)}\right]^{\frac{3}{2}} w_{2}^{4}{\left(t \right)}}\right)=0,&
\end{eqnarray}
\begin{equation}\label{eq:chirp_1}
\frac{d}{d t} w_{1}{\left(t \right)}=2 b_{1}{\left(t \right)} w_{1}{\left(t \right)},  
\end{equation}
\begin{equation}\label{eq:chirp_2}
\frac{d}{d t} w_{2}{\left(t \right)}=2 b_{2}{\left(t \right)} w_{2}{\left(t \right)},  
\end{equation}
respectively. 
The following coupled differential equations for the widths can be derived from Eqs.~(\ref{eq:width_1})-(\ref{eq:chirp_2}): 
\begin{eqnarray}\label{eq:w1_tt}
&\frac{d^2w_1}{dt^2} = \frac{1}{w_{1}^{3}} + \frac{ N_{u} g_{11}}{\sqrt{2\pi} w_{1}^{2}} + \frac{2 N_{v} \gamma_{0} w_{1}}{\sqrt{\pi} \left(w_{1}^{2} + w_{2}^{2}\right)^{\frac{3}{2}}}& \nonumber \\ 
& +\frac{2\sigma^2}{\pi} \left(\frac{3.0 N_{u} N_{v} w_{2}^{2}}{\left(w_{1}^{2} + 2 w_{2}^{2}\right)^{2.5} w_{1}^{2}} + \frac{3 N_{u} N_{v} w_{2}^{2}}{\left(w_{1}^{2} + 2 w_{2}^{2}\right)^{1.5} w_{1}^{4}} + \frac{6.0 N_{v}^{2} w_{1}^{3}}{\left(2 w_{1}^{2} + w_{2}^{2}\right)^{2.5} w_{2}^{3}} - \frac{2 N_{v}^{2} w_{1}}{\left(2 w_{1}^{2} + w_{2}^{2}\right)^{1.5} w_{2}^{3}}\right),&
\end{eqnarray}
\begin{eqnarray}\label{eq:w2_tt}
&\frac{d^2w_2}{dt^2}= \frac{1}{w_{2}^{3}} + \frac{ N_{v} g_{22}}{\sqrt{2\pi} w_{2}^{2}} + \frac{2 N_{u} \gamma_{0} w_{2}}{\sqrt{\pi} \left(w_{2}^{2} + w_{1}^{2}\right)^{\frac{3}{2}}}& \nonumber \\ 
& +\frac{2\sigma^2}{\pi} \left(\frac{3.0 N_{u} N_{v} w_{1}^{2}}{\left(w_{2}^{2} + 2 w_{1}^{2}\right)^{2.5} w_{2}^{2}} + \frac{3 N_{u} N_{v} w_{1}^{2}}{ \left(w_{2}^{2} + 2 w_{1}^{2}\right)^{1.5} w_{2}^{4}} + \frac{6.0 N_{u}^{2} w_{2}^{3}}{ \left(2 w_{2}^{2} + w_{1}^{2}\right)^{2.5} w_{1}^{3}} - \frac{2 N_{u}^{2} w_{2}}{\left(2 w_{2}^{2} + w_{1}^{2}\right)^{1.5} w_{1}^{3}}\right).&
\end{eqnarray}
By identifying the fixed points of this system, one can determine the stationary waveforms.
The symmetric configuration with $w = w_1 = w_2$, $N = N_u = N_v$, and $g = g_{11} = g_{22}$, is more suitable for analytical analysis, and we now carry out the analysis for this simplified case. In this scenario, the system described by equations~(\ref{eq:w1_tt})-(\ref{eq:w2_tt}) can be simplified into the single equation
\begin{equation}\label{eq:w_tt}
\frac{d^2w}{dt^2}= \frac{1}{w^{3}} + \frac{ N (g+\gamma_0)}{\sqrt{2\pi} w^{2}} + \frac{8\sigma^2 N^2}{3\sqrt{3}\pi w^5}.
\end{equation}
One can determine the width of the stationary localized state by solving the algebraic equation
\begin{equation}\label{eq:algebraic_equation}
aw^3+w^2+b=0, \,\,\,\, a = \frac{N(g+\gamma_0)}{\sqrt{2\pi}}; \,\,\, b = \frac{8\sigma^2N^2}{3\sqrt{3}\pi}.
\end{equation}
The real and positive root of this equation is
\begin{equation}\label{eq:roots}
w = -\frac{1}{3a}\left(1+C+\frac{1}{C}\right), \,\,\,\, C = \sqrt[3]{\frac{\Delta_1 + \sqrt{\Delta_1^2-4}}{2}}, \,\,\, \Delta_1 = 2+27a^2b,
\end{equation}
which correspond to the possible value of $w$ that yields a physically valid width for the state, and an amplitude $A = \sqrt{N/(\sqrt{\pi}w)}$. 
\begin{figure}[H]
\begin{center}
\includegraphics[scale=0.32]{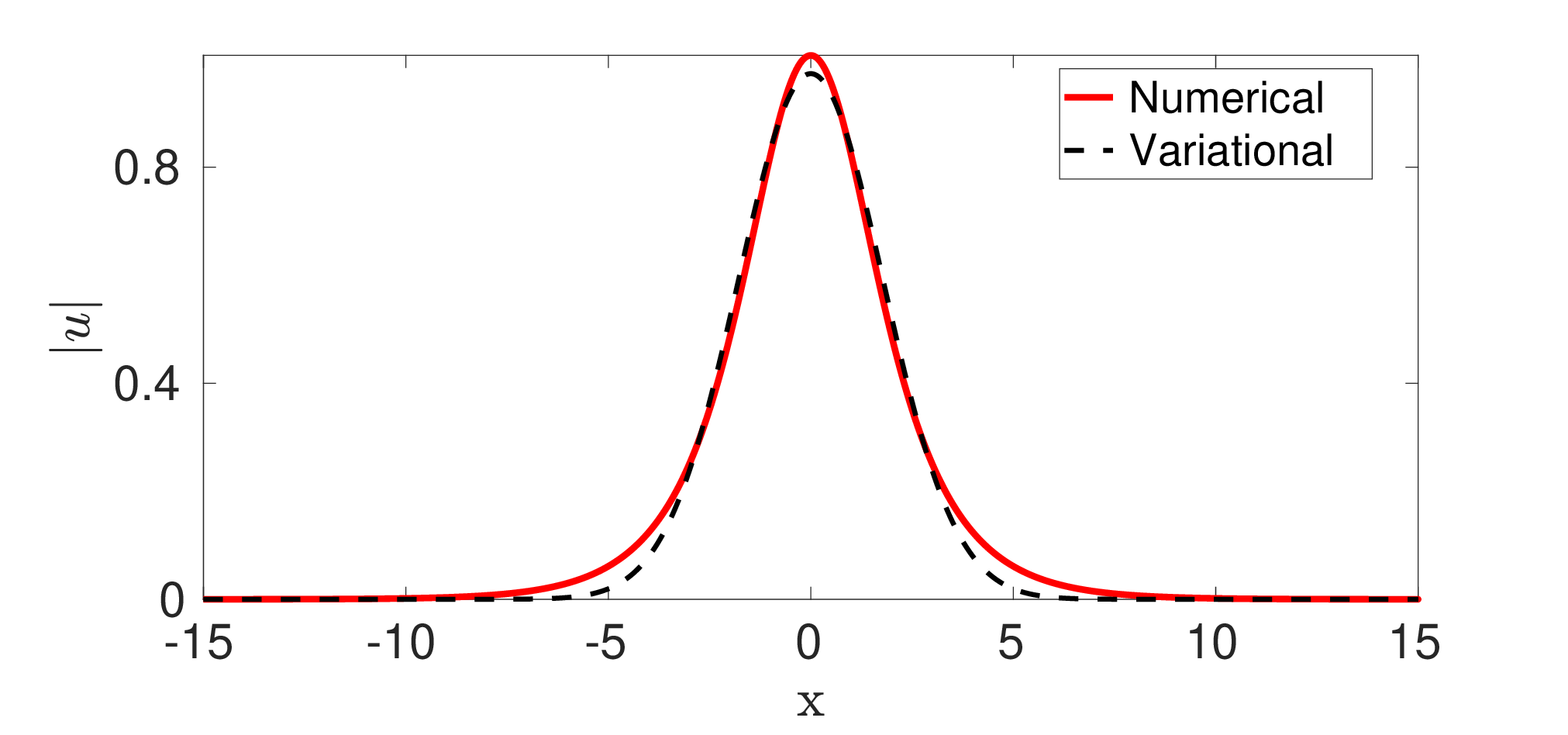} 
\caption{Stationary solutions of the averaged model (\ref{eq:averaged_model}) with parameters: $g=1.5$, $\gamma_0=-2$, $\gamma_1=10$, and $\omega=10 \pi$. Norms of the corresponding components are taken as $N=3$.}
\label{fig:9}
\end{center}
\end{figure}
In Fig.~\ref{fig:9}, comparisons of the stationary solution found by Eq.~(\ref{eq:roots}), and a numerically calculated solution is shown in the case of $N=3$, $g=1.5$, $\gamma_0=-2$, $\gamma_1=10$, and $\omega=10 \pi$. 
For experiments with 39K spin mixtures BEC, see for example~\cite{dErrico, Semeghini}, we have for the parameters $\gamma_1=10, \omega=10\pi$, the scattering lengths 
$a_{11}=a_{22} \sim 20 a_0, a_{12} \sim 200a_0$ and $\omega \sim 10^3$Hz when the trapping frequency is $\omega_{\perp}\sim 10^2$Hz. 
Dynamical evolution of the stationary solutions corresponding to Fig.~\ref{fig:9} are presented in Fig.~\ref{fig:10}. 
We find that the solution becomes unstable over time, such type of instability for the scalar NM model has been observed in~\cite{Itin, Beheshti}.
For other values of the norms ($N$), comparisons of widths and amplitudes found by the variational and the numerical methods are shown in Fig.~\ref{fig:11}. The results shows good agreement for the variational and the numerical analysis.
\begin{figure}[H]
\begin{center} 
\includegraphics[scale=0.35]{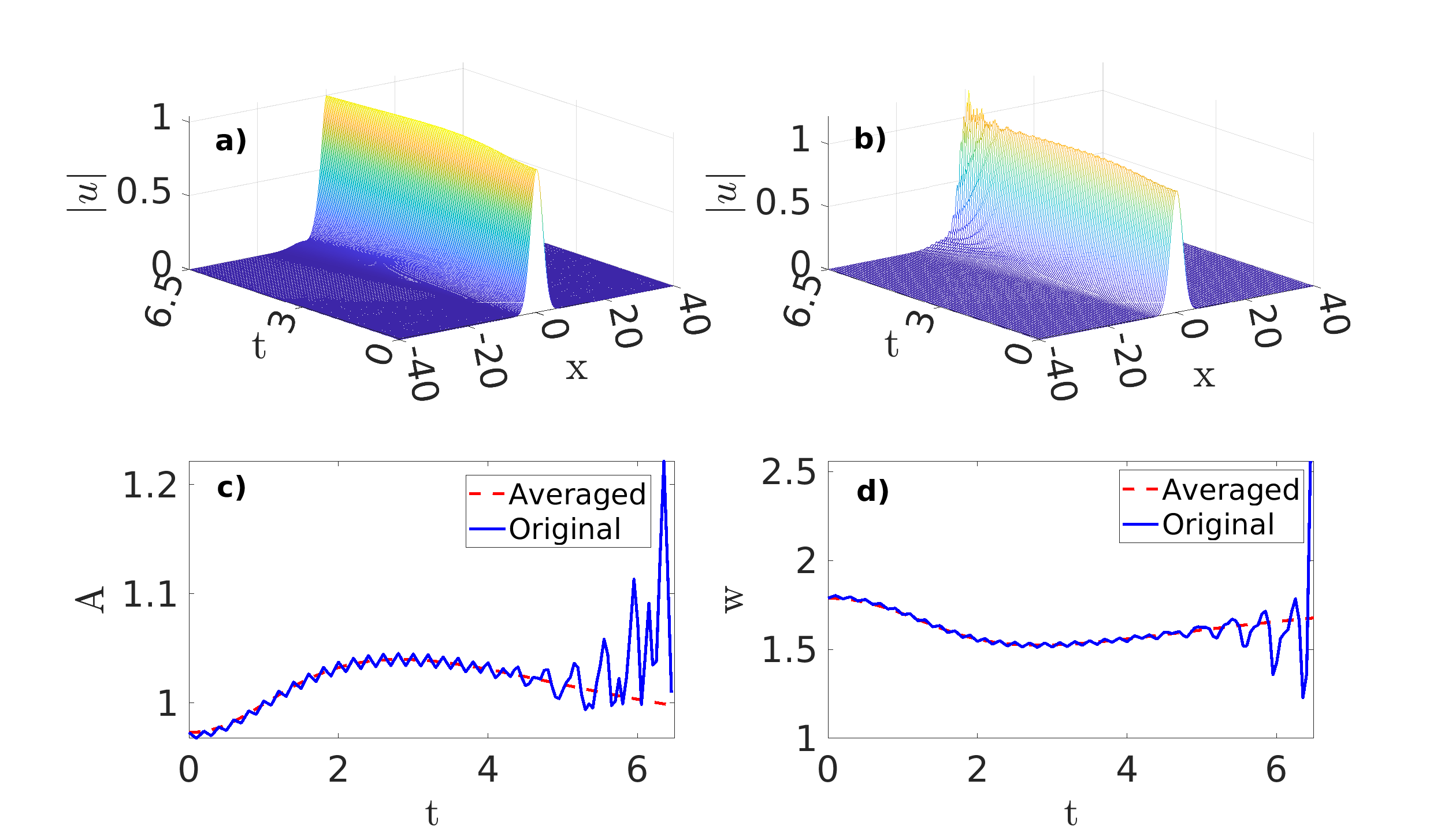} 
\caption{Dynamical evolution of the stationary solutions found by the variational approach. Frame \textbf{a}) represents the numerical solution of the averaged model of Eq.~(\ref{eq:averaged_model}). Frame \textbf{b}) corresponds to the numerical solution of the original model of Eq.~(\ref{eq:original_model}). 
Frames \textbf{c}) and \textbf{d}) corresponds to the amplitudes and widths of the solitons, respectively. In both cases the stationary solutions depicted in Fig.~\ref{fig:9} are used as initial conditions. 
Especially noteworthy is the observation in the comparison that the averaged equation~(\ref{eq:averaged_model}) starts to deviate from the dynamics observed over long time-scales here. 
Parameters used are: $g=1.5$, $\gamma_0=-2$, $\gamma_1=10$, and $\omega=10 \pi$. 
The norms were both taken as $N = 3$.}
\label{fig:10}
\end{center}
\end{figure}

One can also approximate the bound states of the original model~(\ref{eq:original_model}) using asymptotic approximations based on the solution of the averaged model~(\ref{eq:averaged_model}), as discussed in~\cite{ZharPel}. 
For the sake of analytical simplicity, we now consider the symmetric case, which transforms the averaged model into the following conventional NLSE with modified coefficients:
\begin{equation}\label{eq:symmetric_model}
iu_t + \frac{1}{2}u_{xx} - (g+\gamma_0)|u|^2u + \frac{\sigma^2}{2}\left[\left(|u|^2_x \right)^2 + 2|u|^2|u|^2_{xx}\right]u=0.
\end{equation}
To approximate the bound states, we employ a standard ansatz for the stationary solutions: $u(x, t)= \phi(x)e^{-i\mu t}$. 
This leads to an ordinary differential equation (ODE) for the stationary solution $\phi(x)$:
\begin{equation}\label{eq:ODE}
-\frac{1}{2}\phi_{xx} - \mu \phi + (g+\gamma_0)\phi^3 - 2\sigma^2\left(2\phi^3 (\phi_x)^2+\phi^4\phi_{xx} \right)=0.
\end{equation}
The first integral of this equation is given by:
\begin{equation}\label{eq:first_integral}
E = -\frac{1}{2}(\phi_{x})^2 - \mu \phi^2 + \frac{g+\gamma_0}{2}\phi^4 - 2\sigma^2\phi^4(\phi_x)^2.
\end{equation}
Using the same procedure as described in \cite{ZharPel}, it is possible to determine the existence of stationary bright solitons $\phi(x)$ in the open quadrant where $\mu < 0$ and $g+\gamma_0 < 0$. These solitons can be obtained by solving the equation:
\begin{equation}\label{eq:Gaussian_quadrature}
(\phi_x)^2=\frac{(g+\gamma_0)\phi^2-2\mu}{1+4\sigma^2\phi^4}\phi^2,
\end{equation}
with an amplitude of 
\begin{equation}\label{eq:amplitude_analytical}
A = \max \phi(x) = \sqrt{\frac{2\mu}{g+\gamma_0}}.
\end{equation}

In Fig.~\ref{fig:11} we compare the above analytic amplitude with the variational and numerical results.

\section{Extended numerical simulations}
\begin{figure}[H]
\begin{center}
\includegraphics[scale=0.24]{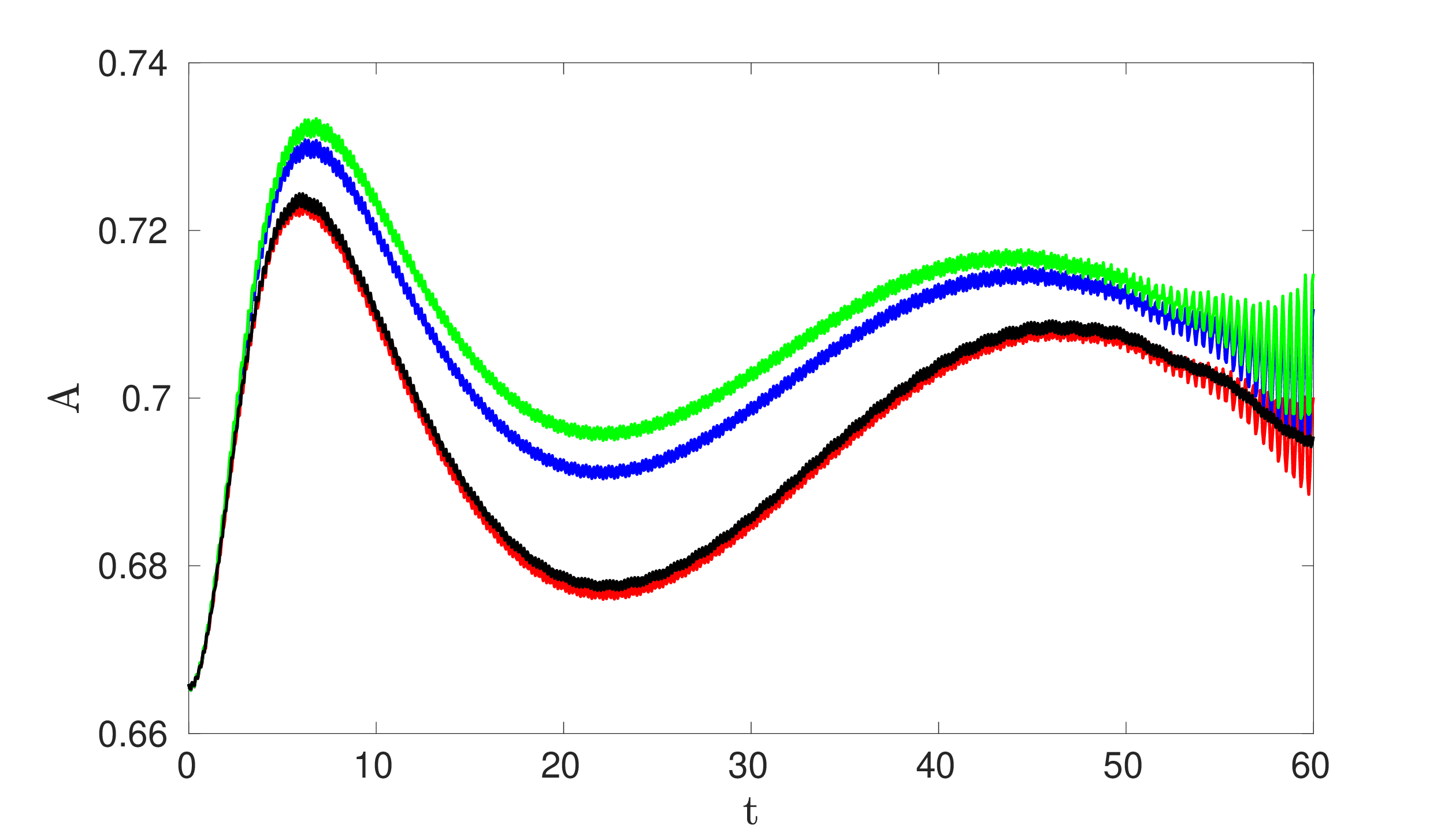}
\includegraphics[scale=0.24]{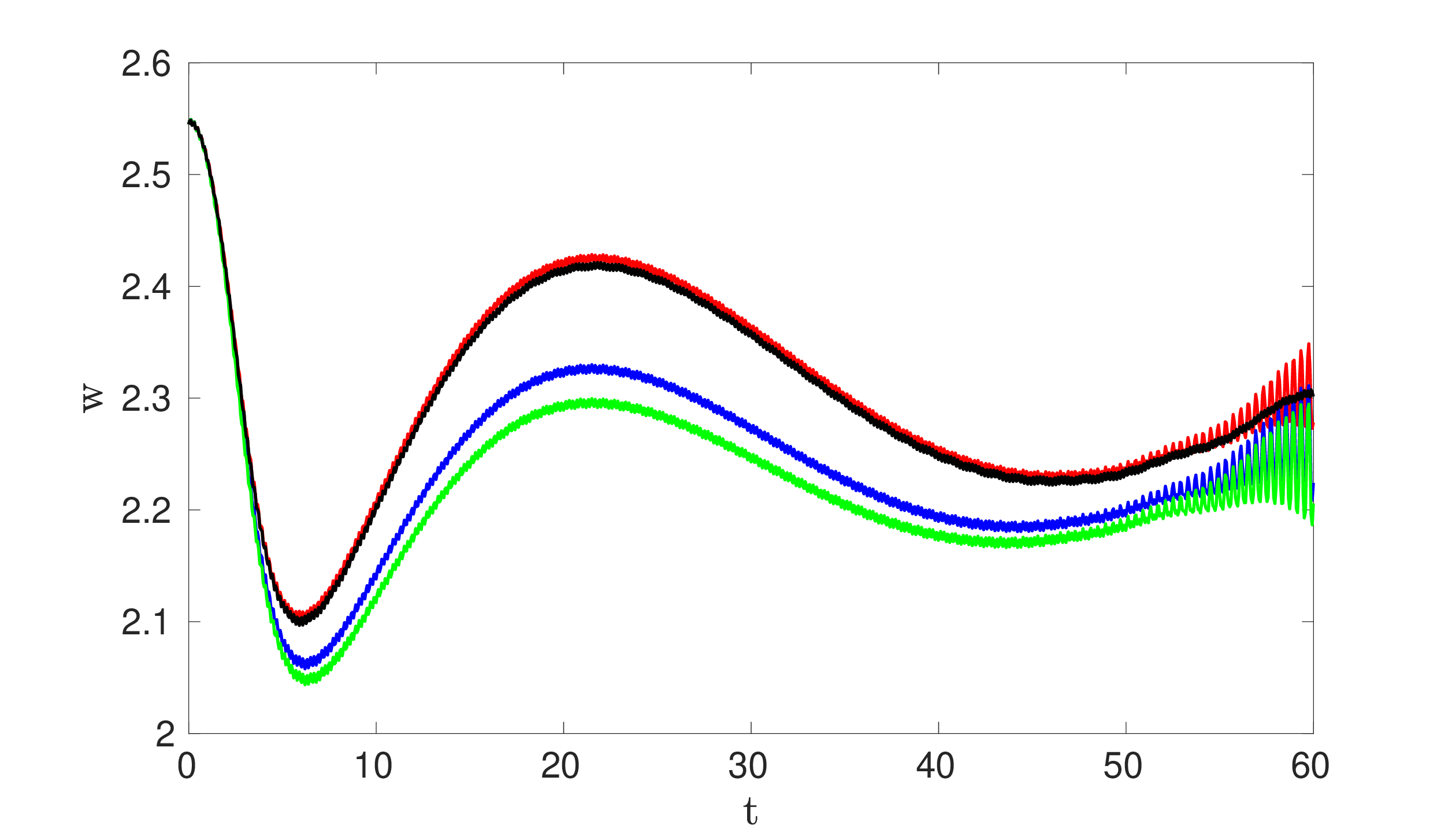} 
\caption{Evolution of amplitude ($A$) and width ($w$) of stationary solutions for $N=2$ with variations in the parameters. The red curve, derived from the variational analysis, see Eq.~(\ref{eq:roots}), reflects the soliton's amplitude and width under the parameters $g=1.5$, $\gamma_0=-2$, $\gamma_1=10$, and $\omega=10\pi$. The other curves demonstrates how altering one parameter, while keeping the rest unchanged, affects the width and amplitude: blue for a slight change in the intra-species interaction ($g=1.485$), green for a minor change in the inter-species interaction ($\gamma_0=-2.02$), and black for a shift in the modulation frequency ($\omega=33$).}
\label{fig:param_change}
\end{center}
\end{figure}

We have also examined how the dynamics of soliton solutions, obtained for specific parameters, respond to small changes in those parameters. As an illustration, Fig.~\ref{fig:param_change} demonstrates the solution dynamics for the case of $N=2$ when the selected parameters are slightly modified. 
In Fig.~\ref{fig:param_change}, the red curve represents the evolution of the amplitude and width of the soliton, as obtained from the variational analysis, see Eq.~(\ref{eq:roots}), with the following parameter values: $g=1.5$, $\gamma_0=-2$, $\gamma_1=10$, and $\omega=10\pi$. 
The other curves depict how the width and amplitude evolve when only one of these parameters is slightly modified, while keeping the remaining parameters constant.

\begin{figure}[H]
\begin{center} 
\includegraphics[scale=0.25]{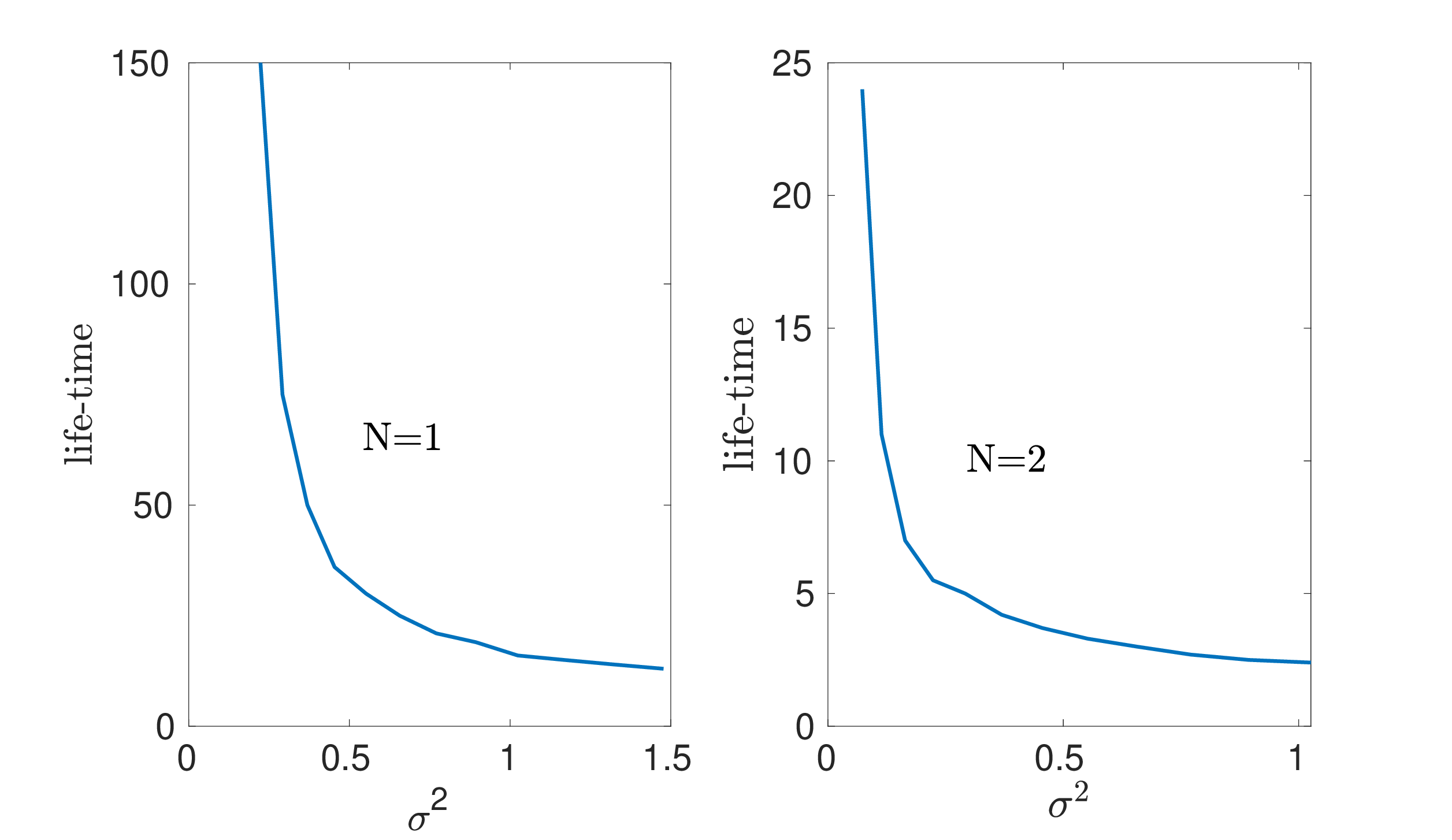}
\includegraphics[scale=0.25]{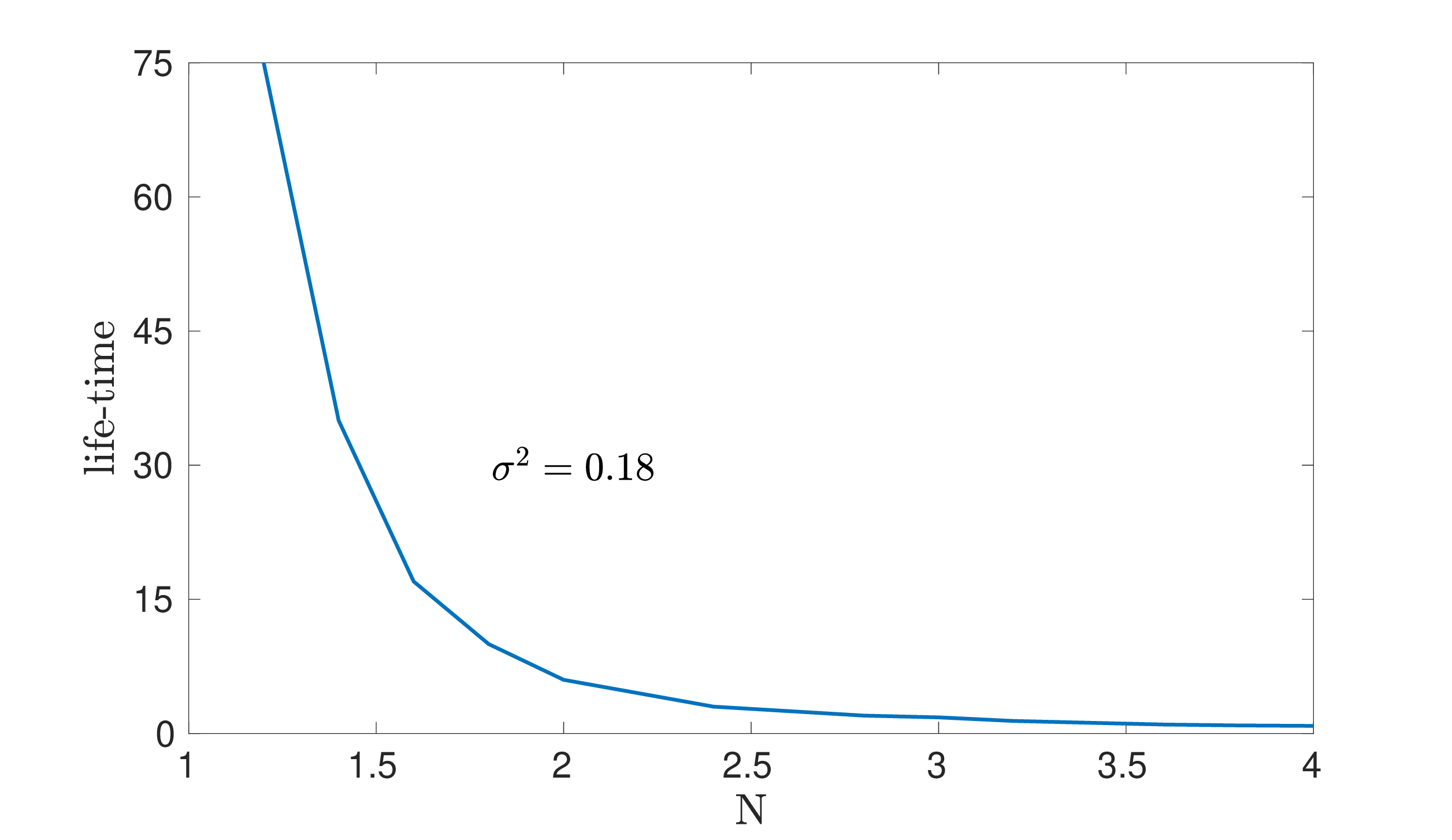}
\caption{Relations of the soliton \textit{lifetime} on the modulation strength ($\sigma^2$) under the norms $N=1$, and $N=2$ (upper two subfigures). 
The relation to the variation of the norm ($N$) at $\sigma^2=0.18$ (lower subfigure).}
\label{fig:7}
\end{center}
\end{figure}

Furthermore, the comparative Fig.~\ref{fig:10} revealed a specific limit to the soliton's \textit{lifetime}, beyond which the averaged equation diverges from the observed long time-scale dynamics.
Extended numerical simulations, carried out over longer durations, revealed a strong dependence of the soliton's \textit{lifetime} on both its norm ($N$), and on the modulation strength ($\sigma^2$). 
Figure~\ref{fig:7} displays the relationship between the soliton's \textit{lifetime} and the modulation parameter strength ($\sigma^2$) under constant norms (top two subfigures), and the relationship between the soliton's \textit{lifetime} and the norm ($N$) with a fixed modulation strength (bottom subfigure).

We have numerically found specific threshold values for both the norm and the modulation strength. 
Below these threshold values, the \textit{lifetime} remains independent of both the norm and the modulation strength. 
To illustrate, in the case of $N=1$ (as shown in the top left subfigure in Fig.~\ref{fig:7}), this threshold value is $\sigma^2 \approx 0.146$ ($\gamma_1=17$). 
In the case of $N=2$ (as depicted in the top right subfigure in Fig.~\ref{fig:7}), this value is $\sigma^2 \approx 0.033$ ($\gamma_1=8$). 
At last, in the case when the modulation strength is $\sigma^2 = 0.18$ (as indicated in the lower subfigure in Fig.~\ref{fig:7}), the threshold value for the norm is $N \approx 0.9$.  
\begin{figure}[H]
\begin{center} 
\includegraphics[scale=0.3]{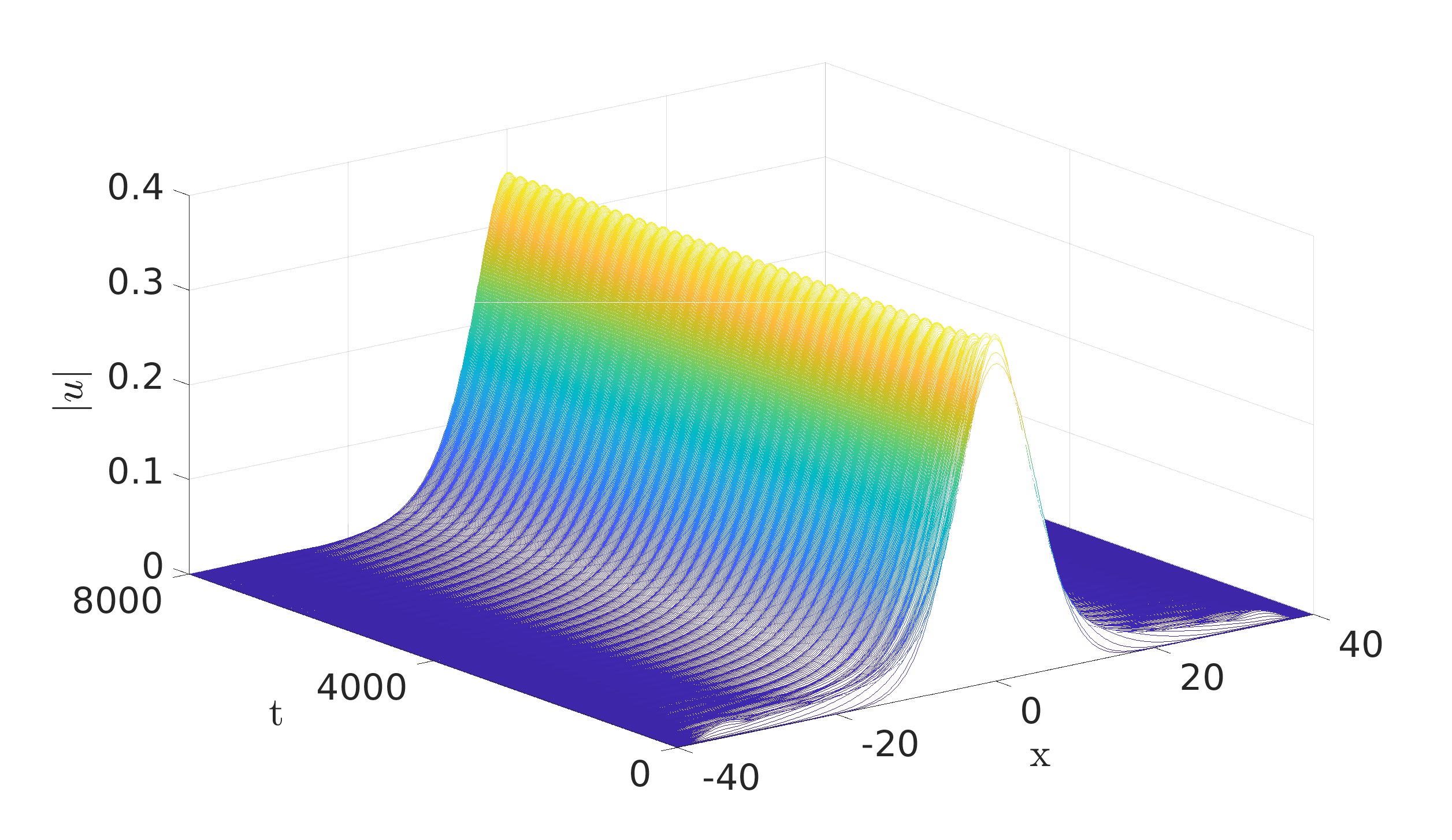}
\hspace{-2mm}~\includegraphics[scale=0.18]{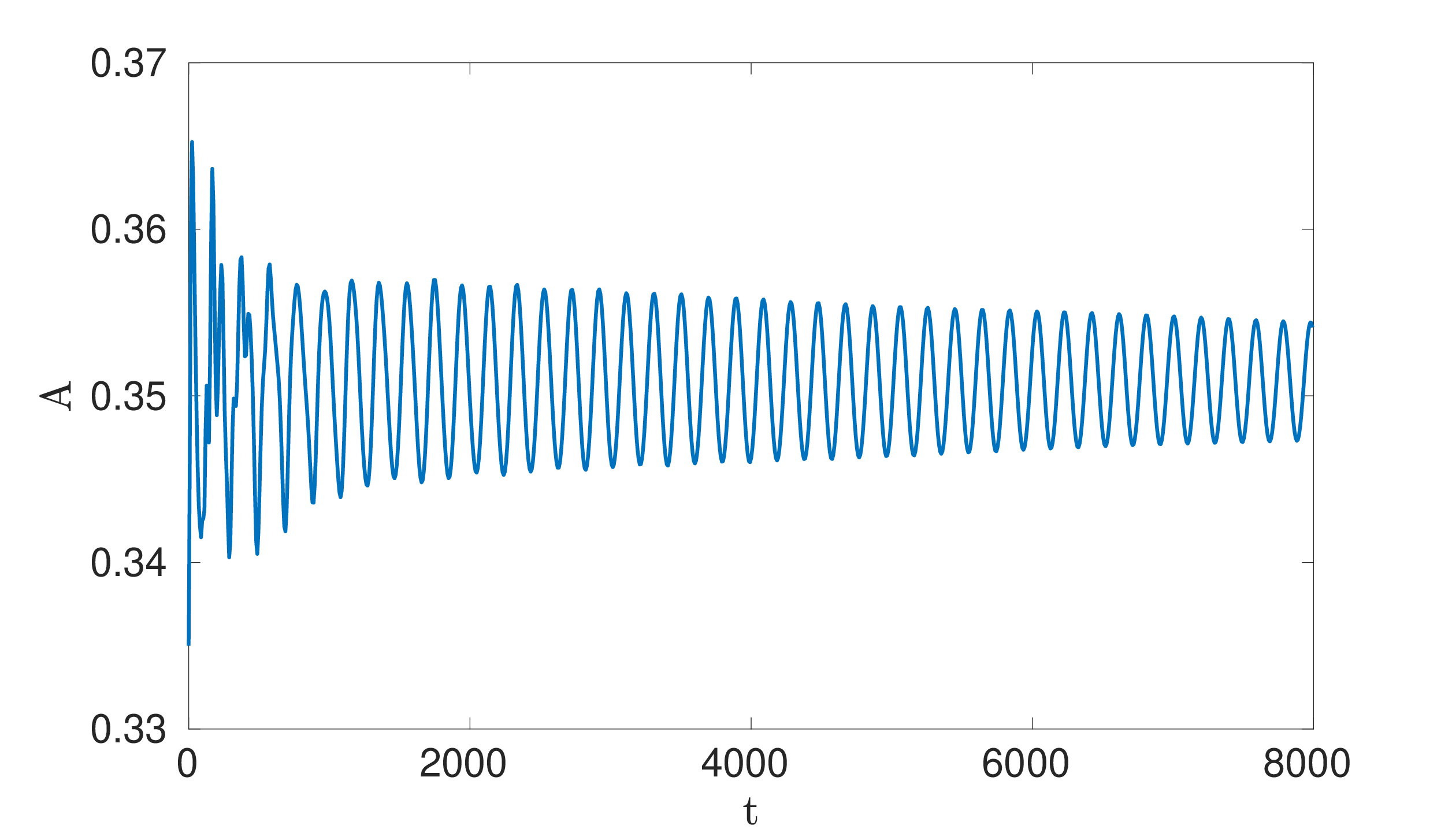}~\hspace{-7mm}~\includegraphics[scale=0.18]{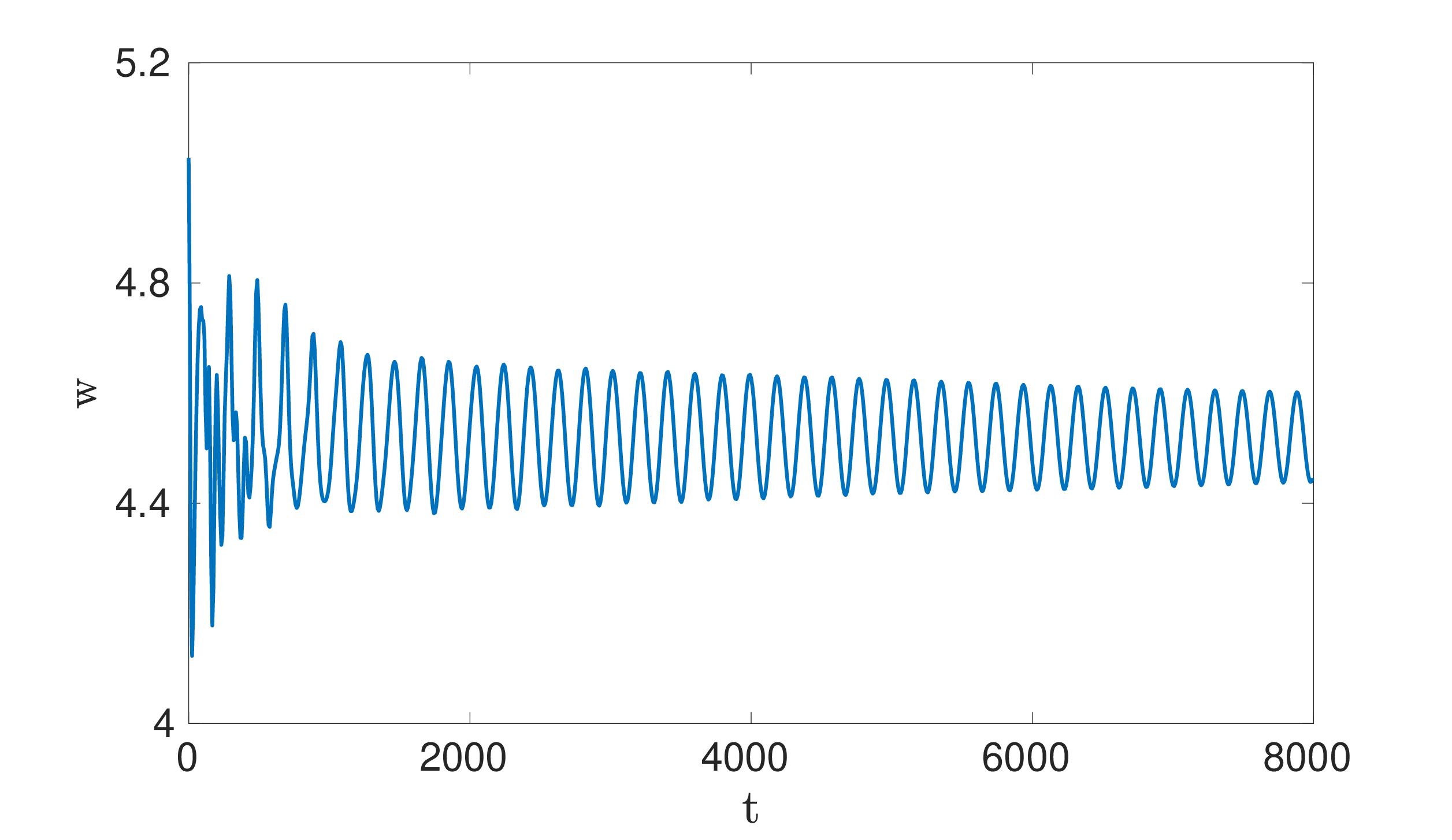}
\caption{Illustration of the long-term evolution of a soliton with below the threshold parameter values: $N=1$, $g=1.5$, $\gamma_0=-2$, $\gamma_1=17$, and $\omega=10 \pi$.}
\label{fig:threshold}
\end{center}
\end{figure}
As an illustration, in Fig.~\ref{fig:threshold}, the behavior of the soliton is depicted when the parameters reach their threshold values. 
It is evident from this figure that the soliton remains intact over a long time-scale without breaking apart.

\section{Conclusions}
In conclusion, our investigation has provided several important results: The first result is the derivation of a system of equations obtained by averaging over rapid and strong modulations of the GPE. The second result reveals that the averaged system exhibits the presence of effective nonlinear quantum pressure, which is more pronounced compared to an unmodulated system. 
Furthermore, the process of modulation instability (MI) has been analytically investigated using the averaged system, and the results are confirmed by numerical simulations of the full time-dependent coupled GPE. 
Moreover, it is shown that the combined action of the linear and nonlinear dispersion from one side, and the mean-field nonlinearities from the other side, leads to the existence of nonlinearity managed (NM) vector solitons. 
Theoretical analysis based on the variational approach was performed, and the predictions of the theory regarding soliton parameters were confirmed through numerical simulations of the full system,
including the sensitivity to variations in the parameter values, and the lifetime of the solitons.

\section*{Acknowledgment}
JSY acknowledges the hospitality of \"Orebro University during a research visit.


\begin{thebibliography}{00}

\bibitem{Malomed1}
B. A. Malomed, {\it Soliton Management in Periodic Systems} (Springer: New York, 2006).

\bibitem{Malomed2}
E. Kengne, W. M. Liu, B. A. Malomed, Phys. Rep. {\bf 899}, 1 (2021).

\bibitem{Abd1}
F. Kh. Abdullaev, J. G. Caputo, R. A. Kraenkel and B. A.
Malomed, Phys. Rev. A {\bf 67}, 013605 (2003).

\bibitem{SU}
H. Saito and M. Ueda, Phys. Rev. Lett. {\bf 90}, 040403 (2003).

\bibitem{Mont1}
G. D. Montesinos, V. M. Perez-Garcia and P. J. Torres, Physica D: Nonlinear Phenomena {\bf 191}, 193-210 (2004).

\bibitem{Mont2}
G. D. Montesinos, V. M. Perez-Garcia, and H. Michinel, Phys. Rev. Lett. {\bf 92}, 133901 (2004).

\bibitem{MalomedOS}
O. V. Matusevich, V. A. Trofimov, E. A. Yudina, and B. A. Malomed, Opt. Spectroscopy, {\bf 106}, 99 (2009).

\bibitem{Abd_pre}
F. Kh. Abdullaev, M. \"Ogren, and J. S. Yuldashev, Phys. Rev. E {\bf 104}, 024222 (2021).

\bibitem{Abd_opt}
F. Kh. Abdullaev, J. S. Yuldashev, and M. \"Ogren, Optik,  {\bf 274}, 170545 (2023). 

\bibitem{Sal1}
M.  Salerno, V. V. Konotop, and Yu. V. Bludov,  Phys. Rev. Lett. {\bf 101}, 030405 (2008).

\bibitem{Nicolin}
A. Balaz, A. I. Nicolin,  Phys. Rev. A {\bf 85}, 023613 (2012).

\bibitem{Sal2}
F. Kh. Abdullaev, P. G. Kevrekidis, and M.  Salerno, Phys. Rev. Lett. {\bf 105}, 113901 (2010).

\bibitem{ZharPel}
V. Zharnitsky and D. E. Pelinovsky, Chaos {\bf 15}, 037105 (2005).

\bibitem{Kevrekidis1}
D. E. Pelinovsky, P. G. Kevrekidis, D. J. Frantzeskakis, and V. Zharnitsky,
Phys. Rev. E {\bf 70}, 047604 (2004).

\bibitem{Kevrekidis2}
P. G. Kevrekidis, D. E. Pelinovsky, and A. Stefanov, J. Phys. A {\bf 39}, 479 (2006);
F.Kh. Abdullaev and J. Garnier, Phys. Rev. E {\bf 72}, 035603(R) (2005). 

\bibitem{Anderson}
D. Anderson, Phys. Rev. A {\bf 27}, 3135 (1983).

\bibitem{dErrico} 
C. d'Errico, M. Zaccanti, M. Fattori, G. Roati, M. Inguscio, G. Modugno, A. Simoni, New Journal of Physics, {\bf 9}, 223 (2007).
%
\bibitem{Semeghini} 
G. Semeghini, G. Ferioli, L. Masi, C. Mazzinghi, L. Wolswijk, F. Minardi, M. Modugno, G. Modugno, M. Inguscio, M. Fattori, Phys. Rev. Lett. {\bf 120}, 235301 (2018).
%
\bibitem{Itin}
A. Itin, T. Morishita, and S. Watanabe, Phys. Rev. A {\bf 74}, 033613 (2006).
%
\bibitem{Beheshti}
S. Beheshti, K. J. H. Law, P. G. Kevrekidis, and M. A. Porter, Phys. Rev. A {\bf 78}, 025805 (2008).
%

\end{thebibliography}
\end{document}